\documentclass[hidelinks]{article}
\usepackage[a4paper, margin = 0.75in]{geometry}
\usepackage{blindtext}
\usepackage{authblk}
\usepackage{graphicx}
\usepackage{setspace}
\usepackage{amsmath}
\usepackage{multirow}
\usepackage{booktabs}
\usepackage{url}
\usepackage{doi}
\usepackage[subpreambles=true]{standalone}
\usepackage{import}
\title{Spatial-temporal dynamics of employment shocks in declining coal mining regions and potentialities of the ‘just transition’}

\author[1,2]{Ebba Mark}
\author[1,2]{Ryan Rafaty} 
\author[1,2,3]{Moritz Schwarz} 

\affil[1]{\small \emph{Climate Econometrics, Nuffield College, University of Oxford}\thanks{We gratefully acknowledge funding from Nuffield College and the Robertson Foundation. We would also like to thank the following people for their insightful contributions: David Hendry, Felix Pretis, Jonas Kurle, and Angela Wenham.}}
\affil[2]{\emph{Institute for New Economic Thinking, Oxford Martin School, University of Oxford}}
\affil[3]{\emph{Smith School of Enterprise and the Environment, University of Oxford}}

\begin{document}
\maketitle

\begin{abstract}
The United States, much like other countries around the world, faces significant obstacles to achieving a rapid decarbonization of its economy. Crucially, decarbonization disproportionately affects the communities that have been historically, politically, and socially embedded in the nation’s fossil fuel production. However, this effect has rarely been quantified in the literature. Using econometric estimation methods that control for unobserved heterogeneity via two-way fixed effects, spatial effects, heterogeneous time trends, and grouped fixed effects, we demonstrate that mine closures induce a significant and consistent contemporaneous rise in the unemployment rate across US counties. A single mine closure can raise a county's unemployment rate by 0.056 percentage points in a given year; this effect is amplified by a factor of four when spatial econometric dynamics are considered. Although this response in the unemployment rate fades within 2-3 years, it has far-reaching effects in its immediate vicinity. Furthermore, we use cluster analysis to build a novel typology of coal counties based on qualities that are thought to facilitate a successful recovery in the face of local industrial decline. The combined findings of the econometric analysis and typology point to the importance of investing in alternative sectors in places with promising levels of economic diversity, retraining job seekers in places with lower levels of educational attainment, providing relocation (or telecommuting) support in rural areas, and subsidizing childcare and after school education programs in places with low female labor force participation due to the gendered division of domestic work.
\end{abstract}

\pagebreak

\tableofcontents

\pagebreak

\setlength{\parskip}{12pt}
\section{Introduction}

The burning of fossil fuels in the United States (US) accounted for 13.5\% of global CO2 energy combustion emissions in 2020, led domestically by oil (43\%), gas (35\%), and coal (19\%), with flaring, cement production and other industrial processes accounting for the final 3\% \cite{data2022a}. Having generated approximately a quarter of all global cumulative carbon dioxide (CO\textsubscript{2}) emissions since 1750, the country’s outsized historical responsibility to decarbonize its economy over the next three decades – consistent with national commitments under the 2015 Paris Agreement – is largely incontrovertible. 
\par
In 2021, President Joe Biden committed his administration to decarbonizing the US economy by decreasing greenhouse gas emissions to 50-52\% below 2005 levels by 2030 with a view towards net-zero emissions by 2050 \cite{whbrief2021}. In August 2022, the Inflation Reduction Act (IRA) was passed constituting a significant step in the direction of these ambitious goals, aiming jointly to reduce inflation, lower prescription medicine prices, and invest heavily in domestic energy production with a focus on clean energy. Climate and energy security investments stipulated in the IRA total 369 billion US\$ \cite{congress2022a}.
\par
As the political momentum behind decarbonization increases, the question of how to usher in a ‘Just Transition’ – ensuring minimal damage from and the equitable sharing of the benefits of decarbonization — is front of mind for environmental and climate justice advocates concerned about impacts on workers, consumers, and frontline communities \cite{newell2013a, mccauley2018a, piggot2019a, pollin2019a, cha2022a}. The IRA makes several long-awaited commitments for consumers and frontline communities, including reducing legacy pollution through block grants, reinstating Superfund taxes to fund toxic site clean-ups and air pollution monitoring, and making clean energy more affordable and accessible for disadvantaged communities. Notwithstanding, environmental justice advocates worry that many of the compromises required to ensure that the IRA was passed, including a guarantee to continue federal oil and gas leasing, could significantly outweigh these benefits \cite{perlin2001a, colman2022a}.
\par
The IRA’s Just Transition provisions for vulnerable workers are more difficult to discern. The International Energy Agency predicts 50,000 layoffs in the US coal sector by 2030 while up to a quarter million US oil and gas jobs are also threatened \cite{iea2021a}. Because fossil fuel deposits are highly concentrated geographically, a decline in output will be felt most sharply in a few select regions with deep roots in the fossil fuel industry \cite{hendrickson2018a, snyder2018a, rickard2020a}. Although the IRA lists several generic initiatives to support American workers, including tax incentives for domestic manufacturing and credits for businesses that pay prevailing wages, only two policy objectives can be described as targeting fossil fuel workers specifically: a 10\% increase in the clean energy tax credit for infrastructure projects operating in or near "energy communities" and a 5 billion US\$ fund providing low-cost loans and refinancing for energy infrastructure, with the notable caveat that qualifying projects should “retool, repower, repurpose, or replace energy infrastructure that has ceased operations” \cite{congress2022a}.
\par
These measures are a just small sample of a larger set of proposed Just Transition policy responses that range from completely compensatory (i.e. direct income support, relocation assistance, interim health care coverage, pension payouts) to active labor market policies (i.e. education, training, and reskilling) and investments in alternative sectors, whether green or otherwise, to provide new job opportunities \cite{pollin2014a, piggot2019a, pollin2019a, pollin2021a}.
\par
To optimize Just Transition aid for at-risk workers, policymakers need to compare available ameliorative measures based on the magnitude, durability, and spatial distribution of adverse employment impacts. While numerous qualitative studies have proffered various conceptualizations of energy justice and ethical principles to guide a ‘Just Transition’ amid structural decarbonization \cite{healy2017a, jenkins2018a}, econometric evidence exploring these real-world barriers to transitional justice is scarce. In-depth regional investigation of the effects of fossil energy industry reliance on employment has generally been done qualitatively, through vulnerability assessments or sociological and ethnographic analyses \cite{bell2010a, abraham2017a, carley2018a, carley2018b, snell2018a, snyder2018a, raimi2021a, sovacool2021a}. The available econometric research has mostly concentrated on oil and gas, specifically oil prices and the macroeconomy \cite{hamilton1983a}. For example, the tight oil and shale gas booms in the early 2000s have garnered the attention of most quantitative studies on employment impacts of fluctuations in fossil fuel production \cite{marchand2012a, weinstein2014a, munasib2015a, paredes2015a, miljkovic2016a}. Furthermore, such results are rarely framed from the Just Transition perspective until a recent study by Scheer et al linked econometric evidence explicitly to the literature through an examination of employment precarity in oil-rich Alberta \cite{scheer2022a}.
\par
We therefore aim to fill this critical evidentiary gap through an in-depth examination of the already experienced, prolonged, and consistent contraction in US coal production. We make three significant contributions to existing literature. We first contribute empirical econometric evidence to elucidate the relevant Just Transition issues, exploring the spatio-temporal nature of coal transitions and employment dynamics in order to illuminate the relative usefulness of proposed policy interventions, including those in the IRA. Second, to the best of our knowledge, ours is also the first panel econometric study to assess the employment impacts of coal mine closures in the US. Most similar to our work, is the examination by Black et al \cite{black2005a} comparing coal and non-coal producing counties in four states during the boom, peak, and fall of the coal industry in the 1970s and 1980s. Other panel studies on US coal production focus mainly on market dynamics and coal productivity or health and safety \cite{stoker2005a, jordan2018a, ghosh2021a}. Third, we establish a typology of coal mining communities to inform future research and policymaking in a period of accelerating energy transitions. To our knowledge, two other studies have defined comparable typologies using theoretical frameworks.\footnote{See Raimi 2021 for a broader review of similar vulnerability assessments.}  Snyder employs three indicators to proxy education, poverty, and rurality, while Raimi proposes twelve \cite{snyder2018a, raimi2021a}. Our work supplements existing literature by presenting an intermediate number of vulnerability axes, striking a balance between multidimensionality and succinctness; by operationalizing the typology through hierarchical clustering analysis; and by reincorporating our typology into our econometric analysis. 
\par
Evaluating the magnitude, persistence, and geographic distribution of employment shocks following mine closures requires a tailor-fit econometric modelling approach that minimizes confounding from unobserved spatial and temporal dependencies. Therefore, we use dynamic panel econometric and spatial econometric methods to assess employment impacts. These methods enable us to account for latent common factors and spatial spill-over effects and achieve greater precision in the estimation of employment shocks caused by structural changes to the US coal sector. We also use machine learning techniques of agglomerative hierarchical clustering to analyse the prominent characteristics of coal mining communities to improve Just Transition policymaking. By integrating clustering analysis with econometrics, we may examine how coal mine closures affect different 'categories' of counties in different ways. Complementarily, we explore the merit of the assumption that green economy investments are sufficient to ensure the stability and prosperity of transitioning fossil fuel producing communities by analysing whether proximate renewable energy investments mitigate unemployment growth following mine closures. The significance of this final question is made all the more relevant given that this assumption underlies one of the main policy provisions of the IRA targeting energy communities.
\par
In what follows, we first describe the data sources and variables used in the econometric and cluster-typological analyses. After discussing our empirical strategy for estimating several panel data models using a suite of purpose-fit specifications, we describe our motivation and method for creating a typology of US coal counties. The study concludes by discussing policy implications and context-dependent policy recommendations based on our combined econometric and cluster-typological analyses.

\section{Data}
\subsection{Econometric Estimation}

First, a panel dataset was created using economic variables from the US Department of Commerce's Bureau of Economic Analysis (BEA), the US Bureau of Labor Statistics (BLS), and the US Census Bureau for 3,072 contiguous US counties from 2002 to 2019. Annual data on active mines in each county came from the Energy Information Administration (EIA) and the Mine Safety and Health Administration (MSHA). The change in the number of active mines\footnote{From the MSHA database, mines were considered active if they had not been classified as “closed” or “abandoned.” This included mines that were labelled as “active” or, in a few cases, “inactive” but not yet closed or abandoned.} in a county in a given year is the main independent variable of interest. The BLS reports the county unemployment rate change from the prior year, our main dependent variable (Table \ref{tbl:models-specs}: Model 1). As dependent variables, we additionally investigate the change in the natural log of employed persons, unemployed persons, total labor force, and population size (Table \ref{tbl:models-specs}: Models 2 to 5). Appropriate combinations of real GDP per capita, real GDP, and population (all in natural log) were included as control variables in each model.
\par
Furthermore, the US Department of Agriculture maintains an energy investment database containing “information regarding USDA programs that provide assistance to renewable energy and efficiency projects” at state, county, and congressional district levels. Since 2002, the USDA has reported a total investment in renewable energy projects of 6.3 trillion US\$ in the contiguous US, or 20.7 US\$ per capita. Only 122.2 million US\$ (7.5 US\$ per capita), or 2\%, of these investments were allocated to coal counties while non-coal counties benefitting from investments received 25.1 US\$ per capita. For the purpose of this study, investments in renewable energy were retained and energy efficiency improvements were removed as they largely represented improvements in on-farm agricultural practices which were deemed unlikely to impact coal workers. The variable of interest derived from this data source is a binary indicator for whether county-level renewable energy investments exceed 0.1\% of county real GDP in a particular year. 
\par
Additionally, the US Census Bureau maintains a county adjacency file for all US counties indicating which counties share a border. This adjacency file was adapted to incorporate information on contiguous US counties when estimating Models 1-5\textsuperscript{[SEM, SLM, SARAR]} as outlined in Table \ref{tbl:models-specs}.

\subsection{Typology}

A dataset of selected social, economic, and political characteristics of the contiguous counties observed in the panel dataset was sourced from the US Census Bureau, US Department of Agriculture, the MIT Election Data and Science Lab, and Chmura, a private producer of economic data that imputes a county-level economic diversity index. Selected variables include educational attainment, population size, median incomes, female labor force participation, economic diversity, rural-urban classification, and political party affiliation.
\par
Appendices A.1 and A.2 of the Supplementary Materials provide details on the indicators, surveys, and data sources used.

\section{Methods}

This study's panel data analysis is based on six baseline model specifications. Table \ref{tbl:models-specs} provides algebraic formulas and technical details of each. All models were estimated using the previously described county-level panel dataset covering 2002 to 2019. Several models (denoted by * in Table \ref{tbl:models-specs}) were estimated on a subset of coal counties, defined as counties that had active mines between 2002 and 2019.  The model specifications are motivated as follows: 

\begin{itemize}
  \item Model 1 evaluates if a change in the number of active mines affects county unemployment rate.
  \item Models 2-5 aim to identify through what channel the unemployment rate varies due to a change in active mines (i.e., a change in the number of employed workers or in the size of the labour force). This is studied by regressing the county unemployment rate determinants (number of employed persons, number of unemployed persons, labour force size, and population size) on the same indicators for changes in active mines.
  \item Model 6 evaluates whether county-level renewable energy investments have lessened the magnitude of employment losses following coal mine closures.
\end{itemize}

Models 1-6 (without superscripts) were estimated using the classic two-way fixed effects (TWFE) estimator, with standard errors clustered by county and year to address within-cluster correlation and heteroskedasticity.\footnote{All TWFE models were estimated using the fixest package in R.} Prior to model estimation, all variables were first-difference transformed to address non-stationarity \cite{sims1990a, castle2019a}. This yielded stationary I(0) covariates, with the additional benefit that the effects of first-difference variables are easy to interpret.
\par

Within this first stream of analysis, two non-negligible issues challenge the interpretation of the results of the Models 1-6: cross-sectional dependence and likely treatment effect heterogeneity.
\par
\begin{table}[p]
\caption{Model Specifications}
\begin{center}
\begin{tabular}{|c|c|c|c|c|} 
 \hline
 \# & \multicolumn{2}{|c|}{Model Specification} & Results \\ 
  \hline
  \rule{0pt}{6ex}\scriptsize{1*}& \multicolumn{2}{|c|}{\parbox{9cm}{\centering\begin{math} \Delta UER_{it} = \boldsymbol{\beta} \begin{pmatrix} \boldsymbol{X}_{it}\\ \Delta log(RealGDPPC)_{it} \\ \end{pmatrix} + \alpha_i + \gamma_t + \varepsilon_{it} \end{math}}} & \parbox{2cm}{\centering \scriptsize Figure \ref{fig:coef_standard}-\ref{fig:coef_all} \\Appendix B.1.2: \\Tables 13-14}\\ [12pt]
 \hline
 \rule{0pt}{4ex}\footnotesize{2*}& \parbox{4.25cm}{\centering\footnotesize\begin{math} \Delta log(Employed Persons)_{it} =\end{math}} & \multirow[c]{3}{*}[0in]{\parbox{6cm}{\centering\begin{math} \boldsymbol{\beta} \begin{pmatrix} \boldsymbol{X}_{it}\\ \Delta log(RealGDP)_{it} \\ \Delta log(Population)_{it} \\ \end{pmatrix} + \alpha_i + \gamma_t + \varepsilon_{it} \end{math}}} & \multirow[c]{4}{*}[0in]{\parbox{2cm}{\centering\scriptsize Figure \ref{fig:uer-decomp} \\Appendix B.1.2: \\Tables 13-14}}\\ [10pt]
 \cline{1-2}
 \rule{0pt}{4ex}\footnotesize{3*}& \parbox{4.25cm}{\centering\footnotesize\begin{math} \Delta log(Unemployed Persons)_{it} =\end{math}} && \\ [10pt] 
 \cline{1-2}
 \rule{0pt}{4ex}\footnotesize{4*}& \parbox{4.25cm}{\centering\footnotesize\begin{math} \Delta log(Labor Force)_{it} =\end{math}} &&\\  [10pt]
 \cline{1-3}
 \rule{0pt}{6ex}\scriptsize{5*}& \multicolumn{2}{|c|}{\parbox{11cm}{\centering\begin{math} \Delta log(Population)_{it} = \boldsymbol{\beta} \begin{pmatrix} \boldsymbol{X}_{it}\\ \Delta log(RealGDP)_{it} \\ \end{pmatrix} + \alpha_i + \gamma_t + \varepsilon_{it} \end{math}}} &\\  [12pt]
 \hline
 \rule{0pt}{6ex}\scriptsize{6*}& \multicolumn{2}{|c|}{\parbox{11cm}{\centering\scriptsize\begin{math} \Delta UER_{it} = \boldsymbol{\beta} \boldsymbol{X}_{it} + \boldsymbol{Z}
 \begin{pmatrix} \Delta Active Mines_{it}*RE Investments_{it}\\ 
 \Delta Active Mines_{it-1}*RE Investments_{it-1}\\ 
 \Delta Active Mines_{it-2}*RE Investments_{it-2} \\
 \end{pmatrix} + \alpha_i + \gamma_t + \varepsilon_{it} \end{math}}} & \parbox{2cm}{\centering \scriptsize Appendix B.1.2: \\Table 19}\\  [18pt]
 \hline
\rule{0pt}{6ex}\scriptsize{1\textsuperscript{SEM}}& \multicolumn{2}{|c|}{\parbox{9cm}{\centering\begin{math} Y_{it} = \boldsymbol{\beta} \boldsymbol{X}_{it} + \alpha_i + \gamma_t + U_{it} \end{math} \\ \begin{math} U_{it} = \delta W_{i}U_{t} + \varepsilon_{it} \end{math}}}  & \multirow[c]{3}{*}[0in]{\parbox{2cm}{\centering\scriptsize Figure \ref{fig:coef_all} \\Appendix B.2: \\Tables 21-23, 25}} \\ [12pt]
 \cline{1-3}
 \rule{0pt}{6ex}\scriptsize{1\textsuperscript{SLM}}&\multicolumn{2}{|c|}{\parbox{9cm}{\centering\begin{math} Y_{it} = \boldsymbol{\beta} \boldsymbol{X}_{it} + \rho W_{i}Y_{t} + \alpha_i + \gamma_t + \varepsilon_{it} \end{math}}} &\\  [12pt]
 \cline{1-3}
\rule{0pt}{6ex}\scriptsize{1-5\textsuperscript{SARAR}}&  \multicolumn{2}{|c|}{\parbox{9cm}{\centering\begin{math} Y_{it} = \boldsymbol{\beta} \boldsymbol{X}_{it} + \rho W_{i}Y_{t} + \alpha_i + \gamma_t + U_{it} \end{math} \\ \begin{math} U_{it} = \delta W_{i}U_{t} + \varepsilon_{it} \end{math}}}  & \\  [12pt]
 \hline
  \rule{0pt}{6ex}\scriptsize{1-6\textsuperscript{HTT(d)}}& \multicolumn{2}{|c|}{\parbox{9cm}{\centering\begin{math} Y_{it} = \boldsymbol{\beta} \boldsymbol{X}_{it} + \sum_{l=1}^{d} \lambda_{il}f(t) + \alpha_i + \gamma_t + \varepsilon_{it} \end{math}}}& \parbox{2cm}{\centering \scriptsize Figure \ref{fig:coef_all}\\ Appendix B.3: \\Tables 26-28}\\  [12pt]
 \hline
 \multicolumn{4}{|c|}{\parbox{15cm}{\small \ \ \\ \textbf{Variable definitions:}\\
\textbf{\textit{i}} : denotes each county for i=1, 2,\ldots3,072 counties\\
 \textbf{\textit{t}} : denotes each year for t=2002, 2003,\ldots2019  \\
 $\boldsymbol\alpha_{i}$ : unit-fixed effects \\
 $\boldsymbol\gamma_{t}$ : time-fixed effects\\
 $\boldsymbol X_{it}$ : Vector of independent variables ($\Delta$ Active Mines\textsubscript{it}, $\Delta$ Active Mines\textsubscript{it-1}, $\Delta$ Active Mines\textsubscript{it-2}). For Model 6, this includes levels of all interacted variables. For Models 1-5\textsuperscript{[SEM, SLM, SARAR]} and 1-6\textsuperscript{HTT(1),HTT(2)]} this represents all independent variables denoted in Models 1-6, respectively.\\
 $\boldsymbol Y_{it}$ : Relevant dependent variable denoted by Model number.\\
 $\boldsymbol\beta$ : Vector of regression coefficients to be estimated.\\
 $\boldsymbol\rho$ : Spatial autoregression parameter.\\
 $\boldsymbol\delta$ : Spatial autocorrelation parameter.\\
$\boldsymbol W_{i}$ : N X N spatial weight matrix for neighbors of unit \textit{i} where N = 3,072.\\
 $\boldsymbol\lambda_{il}$ : Individual loadings parameter per unit \textit{i} and factor \textit{l} in heterogeneous trend models.\\
\textbf{\textit{f(t)}} : Unobserved time-varying common factors used in heterogeneous trend models.\\
\textbf{\textit{REInvestments\textsubscript{it}}}: dummy indicator of whether renewable energy investments exceeded 0.1\% of GDP\textsubscript{it}.\\
$\boldsymbol\Delta UER_{it}$ : Change in unemployment rate in county \textit{i} in year \textit{t}.\\
\textit{Note: * indicates model was also applied to a subset of coal counties only.}}}\\ 
  \hline
\end{tabular}
\label{tbl:models-specs}
\end{center}
\end{table}

\subsection{Accommodating Cross-Sectional Dependence}

The TWFE estimator, like other estimators used in ordinary least squares regression modelling, requires the restrictive condition of cross-sectional independence. In practice, any unobserved cross-section dependencies, such as structural breaks, global stochastic trends, or spatial (spill-over) effects, in the underlying data generating process are assumed to be captured sufficiently using linear additive two-way (unit and time) fixed effects. In many real-world circumstances, this assumption is too restrictive \cite{imai2019a, chaisemartin2020a}. Although it can accommodate time-invariant confounders, the TWFE estimator comes at the expense of potential dynamic misspecification \cite{plumper2019a}. Insufficient accounting for cross-sectional dependence can lead to inaccurate and inconsistent regression coefficient estimates \cite{anselin1998a, pesaran2014a}. And indeed, this concern appears justified in the present study as several diagnostic tests could not reject the presence of significant cross-sectional dependence in Models 1-6 \cite{anselin1988a, anselin1996a, millo2017a}.
\par
Therefore, two estimation approaches frequently used to account for cross-sectional dependence were applied to complement the TWFE OLS model results. Specifically, we use estimators that account for spatially correlated effects (SEM, SLM, and SARAR in Table \ref{tbl:models-specs}) and one- and two-factor heterogeneous trends models (HTT(1) and HTT(2) in Table \ref{tbl:models-specs}). Each method is described below.

\subsubsection{Spatial Econometric Models}

First, three spatial econometric models were estimated: a spatial autoregressive model (SLM), a spatial error model (SEM), and a spatial autoregressive model with autoregressive error structure (SARAR) \cite{lee2010a, baltagi2021a}.\footnote{All spatial models were estimated using the splm package in R.}  The latter combines the SLM and SEM models' core features. Given the extent of spatial dependency in our data \cite{hendry1978a, mizon1995a}, the SLM and SARAR models were selected as the most acceptable spatial feature specifications based on AIC and BIC information criteria.\footnote{Information about testing for residual spatial dependence on spatial models can be found in Appendix C.1 of the Supplementary Materials.}  Beyond serving as a robustness check on our initial TWFE estimates, the SLM and SARAR models provide additional valuable information: namely, they tell us whether the impact of mine closures exhibit regional spill-over effects.

\subsubsection{Heterogeneous Trends Models}

The second method employed to accommodate cross-sectional dependence is via latent factor modelling, which assumes the panel data model exhibits a factor-analytic error structure. In this approach, one or more latent (unobserved) common factors can be estimated via iterative principal component analysis (the "interactive fixed effects" estimator proposed in Bai (2009)), or instead, approximated via cross-sectional averages of the dependent and independent variables (the "common correlated effects" estimator proposed in Pesaran (2006)) \cite{pesaran2006a, bai2009a, kneip2012a}. Another method is to estimate the unobserved factors semi-parametrically via functions that capture unit-specific time trends (the "heterogeneous time trends" estimator proposed in Kneip et al (2012)). We apply the heterogeneous time trends estimator because it permits unobserved components to be non-stationary and exhibit smooth time trends. It is therefore well-suited for capturing unobserved trends related to long run energy prices, automation, and technical efficiency advances.\footnote{In contrast, the method of Bai (2009) assumes the latent factors are stationary (stochastically bounded), and although the method of Pesaran (2006) can be validly extended to settings with non-stationary factors (see: Kapetanios et al. 2011), its small sample properties are not ideal for the present study’s dataset and for the dynamic model specifications we wish to estimate.}  The AIC or BIC criteria for one- and two-factor models showed no significant benefit of including more than one factor when computing heterogeneous time trends.\footnote{These heterogeneous trend models were implemented using the phtt package in R.}
 
\subsection{Accommodating Treatment Effect Heterogeneity}

The TWFE estimator may also generate bias when treatment effects are entirely or partially heterogeneous across units. To properly infer the effect of a treatment variable (e.g., a change in the number of active coal mines) on a target variable (the county unemployment rate), the TWFE estimator forces the econometrician to assume the treatment effect is constant across units and time, known as the "common trends" assumption. In the present study, there are compelling reasons to suspect that employment impacts of mine closures can vary substantially across counties for reasons other than those which we are able to capture via the aforementioned control variables. Specifically, unmodeled demographics can cause selection bias. These demographic traits need not be directly related to changes in the unemployment rate but may predict counties' transitional capacities when mines close. By identifying weighted sums of average treatment effects in each group and period without considering the underlying reason for the variation, the TWFE estimator may not capture relevant treatment effect heterogeneities across units, leading to biased and potentially misleading coefficient estimates \cite{chaisemartin2020a}. 

\subsubsection{Agglomerative Hierarchical Clustering}

Therefore, to address the issue of possible treatment effect heterogeneity in our sample, agglomerative hierarchical clustering analysis was performed on the typology dataset containing information on salient demographic characteristics of the 252 counties identified as coal counties as well as the total set of 3,072 contiguous US counties \cite{ketchen1996a, murtagh2012a, rodriguez2019a}. Prior to running the clustering analysis, the indicators used were scaled or standardized to have a mean of zero and standard deviation of one. Applying three of the most common methods (elbow, silhouette, and gap statistic) for determining the optimal number of clusters to identify yielded inconclusive results for the coal counties, proposing between two and four clusters \cite{rousseeuw1987a, tibshirani2001a}. In the case of the complete contiguous US dataset the three tests to identify the optimal number of clusters revealed consistent results, suggesting three as the optimal number of clusters. Thus, to aid in an eventual comparison of coal counties to US counties overall, each dataset was clustered into three “types.” Variables considered in the clustering analysis include an urban/rural index, population size, educational attainment, economic security, female labor force participation, economic diversity, and political attitudes.
\par
This hierarchical clustering serves a dual purpose. First, the resulting ‘cluster membership’ information obtained from the tripartite typology of coal counties was used to re-estimate Model 1 using grouped fixed effects, allowing for partially heterogeneous slope coefficients for each independent variable of interest.\footnote{The modified Model 1 with grouped fixed effects was estimated using the fixest package in R.} Second, we present a constructed typology that sheds light on the diversity and geographical distribution of various challenges faced by coal communities across the US. 
\par
The results of all aforementioned regressions (Appendices B and D), tests (Appendix C), model selection methods (Appendix B.2), and clustering (Appendices C.3 and D) can be found in the Supplementary Materials.

\section{Results}
\subsection{Two-Way Fixed Effects, Spatial, and Heterogeneous Trends Econometric Models}

When estimating the models presented in Table \ref{tbl:models-specs}, we expect that the closure of a coal mine would spark an increase in the county-level unemployment rate. Such an effect is unlikely to occur solely because of employment losses in the coal industry specifically but likely in combination with spill-over effects from such employment losses onto the jobs and livelihoods of those whose work is conducted in support of coal county functioning (i.e., education, services, retail). This hypothesis is consistent with macroeconomic theory positing that there are direct, indirect, and induced jobs associated with a given economic activity \cite{bacon2011a}.  
\par
\begin{figure}
\centering
\includegraphics[width=10cm]{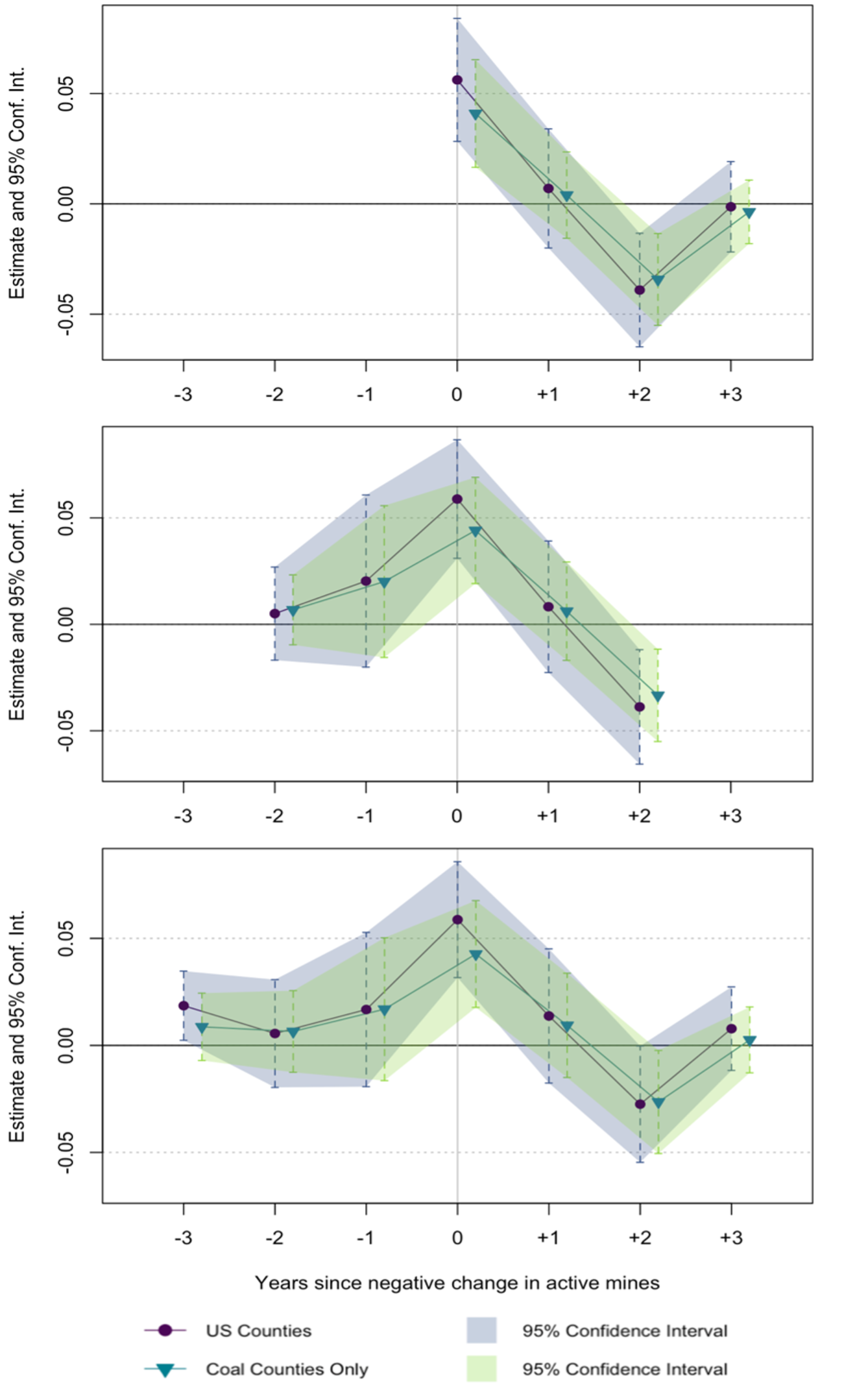}
\caption{\emph{Figure \ref{fig:coef_standard} illustrates the sequence of responses in county unemployment rate to a change in active mines at time t = 0 and associated 95\% confidence intervals. For simplified interpretation, the sign of the impact has been flipped to illustrate the change in unemployment rate resulting from a decrease in active mines. The y-axis indicates the time since the change in active mines was reported. Three combinations of lags and leads have been included for illustrative purposes as they do not significantly change the principal coefficient estimates at times t, t-1, and t-2. Exact coefficients are reported in Tables 12 and 13 of Appendix B.1.2 of the Supplementary Materials.}}
\label{fig:coef_standard}
\end{figure}

\par
Model 1 finds that a single mine closure (one-unit decrease in the number of active mines) is associated with a 0.056 (0.041; restricted coal county sample in parentheses) percentage point increase in county unemployment rate. For the model using all US counties (coal counties only), we reject the null hypothesis of no effect at the 0.1\% (1\%) level. When evaluating dynamics, the unemployment rate neither seems to continue to increase nor recover a year after mine closures are reported with a further change in unemployment rate imperceptibly different from zero. However, two years later, the unemployment rate appears to recover slightly, decreasing by 0.039 (0.034) percentage points at a 1\% (1\%) level of statistical significance on the dataset of all US counties (coal counties). A linear combination of the coefficient estimates confirms the non-persistent nature of the unemployment rate change. Coefficient estimates are reported in Figure \ref{fig:coef_standard} of the main text and Tables 13 and 14 of Appendix B.1.2 of the Supplementary Materials.
\par
Figure \ref{fig:coef_all} compares coefficient estimates reported by Models 1, 1\textsuperscript{[SEM, SLM, SARAR]}, and 1\textsuperscript{[HTT(1), HTT(2)]} models along with their respective 95\% confidence intervals. The coefficients reported in Figure \ref{fig:coef_all} for Model 1\textsuperscript{SLM} and Model 1\textsuperscript{SARAR} are the direct (intra-county) and indirect (inter-county) impacts of a change in active mines in a particular county, reported separately.\footnote{Coefficient interpretation for the SLM and SARAR models requires calculating the direct intra-county and indirect inter-county impacts of the independent variables on the outcome variables (Piras, 2013). These impacts were calculated using the spatialreg package in R.}  The direction of responses in the unemployment rate are consistent across all estimations, providing substantial validation of the sign of the coefficient estimates derived from Models 1-6. Although the SLM and SARAR models estimate the direct impact of mine closures to be slightly smaller in magnitude (significant at the 0.1\% level for a closure in time t and t-2), the indirect and combined total impacts are consistently larger, indicating the presence of strong spatial diffusion of impacts not identifiable in the TWFE model. Model 1\textsuperscript{SARAR}, for example, finds a contemporaneous 0.215 percentage point increase in unemployment rate from a one-unit decrease in active mines in a neighbouring county, 6 times greater than the direct impact identified by the same model. This spatial diffusion echoes results in previous applications of spatial econometric models to US employment data 
 \cite{bronars1987a}.
\par
\begin{figure}[h]
\centering
\includegraphics[width=15cm]{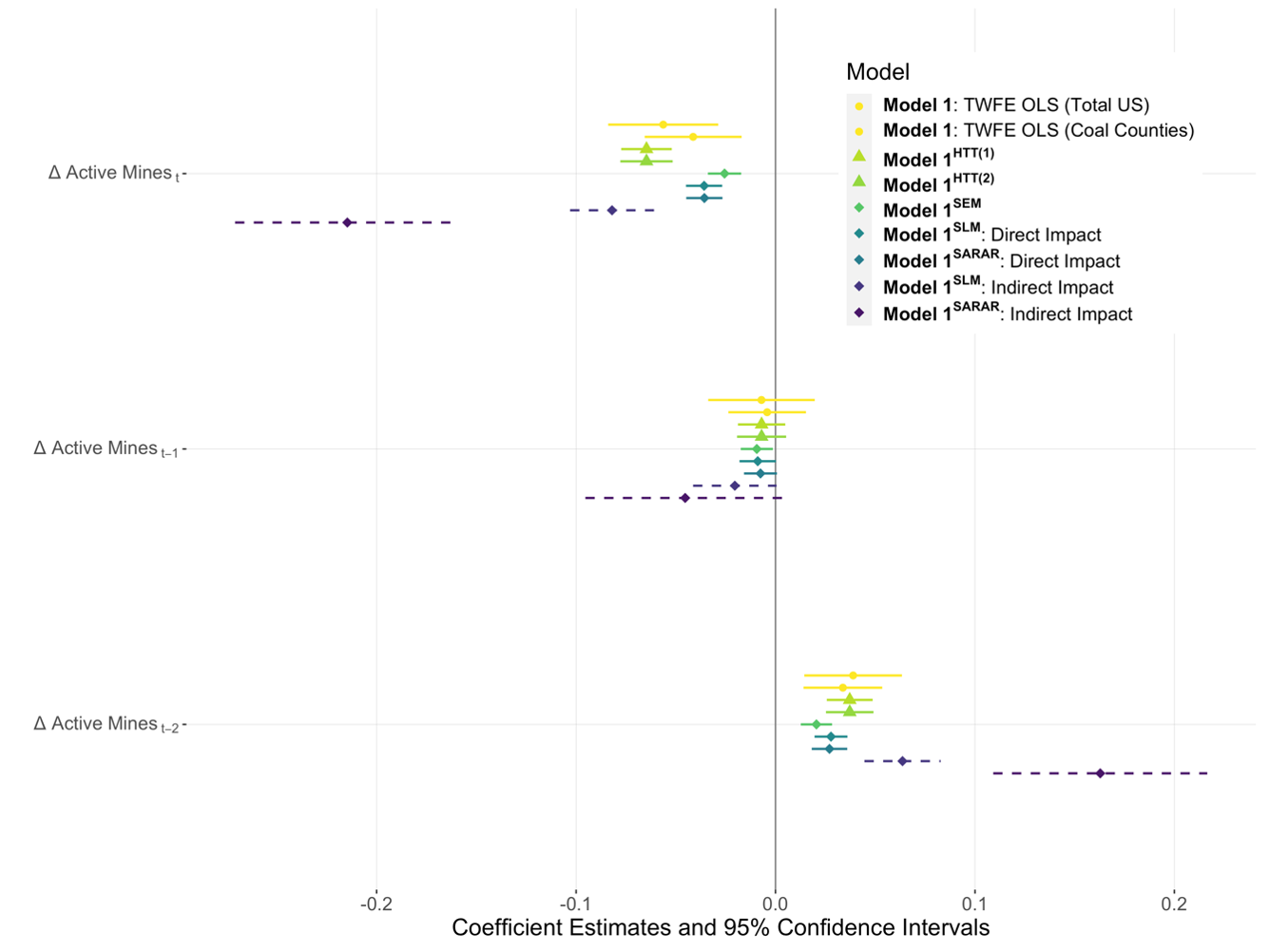}
\caption{\emph{Coefficient estimates and corresponding 95\% confidence intervals represent the response of unemployment rate to a change in active mines in times t, t-1, and t-2 as estimated by each model estimation. Confidence intervals for “indirect” impacts calculated from the SLM and SARAR models are represented by dashed lines.}}
\label{fig:coef_all}
\end{figure}

\par
Similarly, the coefficient estimates of the heterogeneous trends models are greater in the same year in which a change in active mines occurs. More precisely, the one-factor heterogeneous trends model Model 1\textsuperscript{HTT(1)} reports that a one-unit decrease in the number of active mines is associated with a 0.064 percentage point increase in unemployment rate at a 0.1\% level of statistical significance. Given that heterogeneous trends models have a de-biasing effect on coefficient estimates due to the consideration of unobserved latent factors, this result indicates that the response in unemployment rate in the same year as a change in active mines is likely underestimated in Model 1. In short, the spatial and heterogeneous trends models indicate a significant underestimation of the true contemporaneous impact of a mine closure on county unemployment rate. The heterogeneous trends models indicate that the estimate is nearly 1.2 times greater and the spatial models indicate that much of this unemployment rate spike is reflected in neighbouring counties. Coefficient estimates of Models 1\textsuperscript{[SEM, SLM, SARAR]} and Models 1\textsuperscript{[HTT(1), HTT(2)]} are represented in Figure \ref{fig:coef_all} and can be found in Appendices B.2 and B.3, respectively, of the Supplementary Materials.
\par
To add further detail, asymmetric estimation was carried out to evaluate whether increases and decreases of active mines have symmetric effects.  The regression results of each of these estimations is provided in Tables 17 and 18 of Appendix B.1.2 in the Supplementary Materials and show that the magnitude and statistical significance of a response in the unemployment rate to a change in active mines is mainly attributable to a negative change. Linearly combining the coefficient estimates confirms this finding and reveals further illuminating results.  First, when incorporating a factor for whether the change was negative, the magnitude of change in time t = 0 increases in magnitude to 0.075 percentage points (compared to 0.056 percentage points in Model 1) at the 1\% significance level. Second, when incorporating a factor for whether the change in mines was positive, the magnitude of change decreases in magnitude to just –0.019\% and is no longer statistically significant. Altogether, this asymmetric treatment model indicates minimal, if any, employment benefits from counteracting a coal phase-out.

\subsection{Putting It Into Context: Determinants of the Unemployment Rate Response}

\begin{figure}[h]
\centering
\includegraphics[width=15cm]{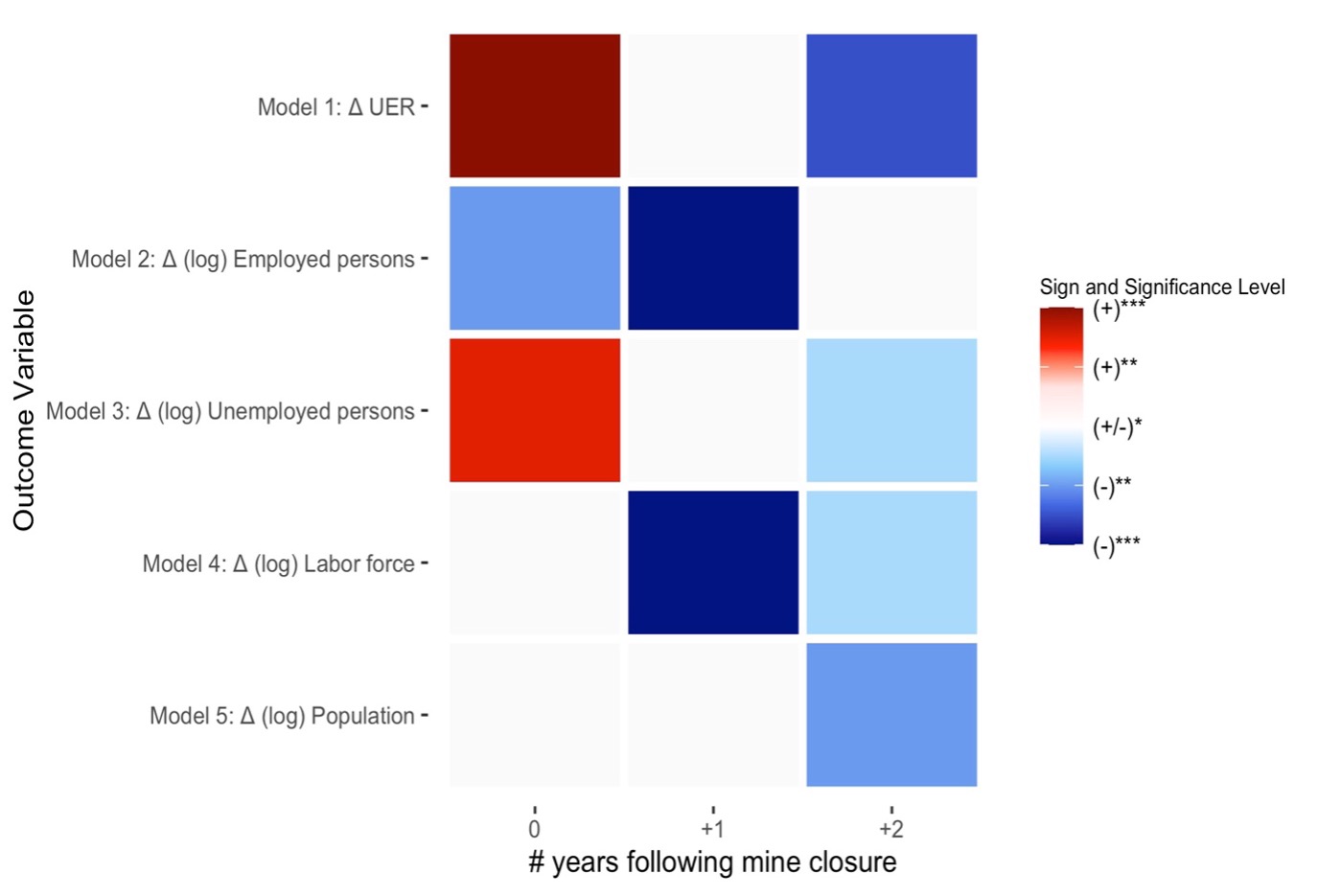}
\caption{\emph{This figure indicates the direction of changes (red = increase; blue = decrease) in the unemployment rate, natural log of employed persons, natural log of unemployed persons, natural log of labor force, and natural log of population at the county level from a change in active mines at time t, t-1, and t-2. The sign of the response variable has been flipped to allow for interpretation of the sign as the response to a decrease in the number of active mines. The sign and magnitude of the response is reported from the model run on all US counties. Significance codes are reported as follows: *** 0.1\%, **1\%; *5\%. A grey box indicates an insignificant variable response.}}
\label{fig:uer-decomp}
\end{figure}

Our initial results do not reveal which determinants of unemployment rate might be causing these responses to mine closures.  To answer this question, we considered the main variables used to calculate the rate: unemployed persons, employed persons, and total labor force (the sum of employed and unemployed persons). Their relationship is defined by Equation \ref{eq:uer-eq} below:

\begin{equation}
Unemployment\:Rate = \frac{Unemployed\:Persons}{Unemployed + Employed\:Persons} = \frac{Unemployed\:Persons}{Labor\:Force}
\label{eq:uer-eq}
\end{equation}
The sign and statistical significance of the change in each variable 0, 1, and 2 years following a mine closure are reported in Figure \ref{fig:uer-decomp}.
\par

First, in the year when the change in active mines is reported, the change in unemployment rate is found to most likely be caused by a reshuffling of the labor force from employed to unemployed persons, or lay-offs, rather than any change in the size of the labor force or population overall. 
\par
In the following year, the unemployment rate neither continues to increase nor recover. The only two detectable changes in the factors of the unemployment rate are further decreases in employed persons and labor force size. This not only indicates that employment continues to decrease but also that those leaving employment are not entering unemployment and therefore likely either become discouraged, retire, or move to another county. However, there is no discernible change in population size, and it is therefore more likely that the shrinking labor force indicates either discouraged or retired workers leaving the category of employed persons. Although the magnitude of change in population size discerned in our study is small, such a theory is consistent with well-documented underemployment challenges in coal regions like Appalachia \cite{wood1960a}.  
\par
Finally, two years following a change in active mines, we observe a small decrease (or partial recovery) in county unemployment rate in Model 1. However, Figure \ref{fig:uer-decomp} suggests that this is not principally determined by a re-entry of unemployed persons into employment. Rather, a further shrinking of the labor force and decrease in unemployed persons considered together with a small decrease in overall population size indicates that workers might be moving to meet other job opportunities. This result echoes findings that local demand shocks induce a stock equilibrium shift through migration such that the relative unemployment rate returns to its equilibrium state \cite{treyz1993a}. Furthermore, past applications of spatial models to regional unemployment rates in the US and UK have indeed found that shocks induce a transitory change in unemployment rate that equilibrates following a regional adjustment process \cite{molho1995a}.
\par
The direction of the coefficients are confirmed by Models 1-5\textsuperscript{SARAR} models and Models 1-5\textsuperscript{HTT(1)} reported in Appendices B.2 and B.3 in the Supplementary Materials.

\subsection{Renewable Energy Investments: Evaluating the Potential of the Inflation Reduction Act}

In Model 6 we find small and statistically insignificant alleviating impacts of renewable energy investments on unemployment rate changes caused by mine closures, while the other results remain robust. Investigating further whether renewable energy investments are incapable of alleviating employment losses caused by decarbonization or whether past green investments were simply too small is beyond the scope of this study. However, regardless of reason, the inability to detect such an impact at this stage is still of significance to the validity of the discourse surrounding the potential of the green economy to offset losses in brown industries. Given one of the only targeted provisions for workers in coal (and other fossil fuel producing) communities in the IRA relies heavily on the assumption that renewable energy investments can shore up workers, further study on this topic would be most relevant. Regression coefficients for Model 6 can be found in Appendix B.1.2 and Model 6\textsuperscript{HTT(1), HTT(2)} in Appendix B.3 in the Supplementary Materials.
\par

\subsection{Typology}

\begin{figure}[h]
\centering
\includegraphics[width=12cm]{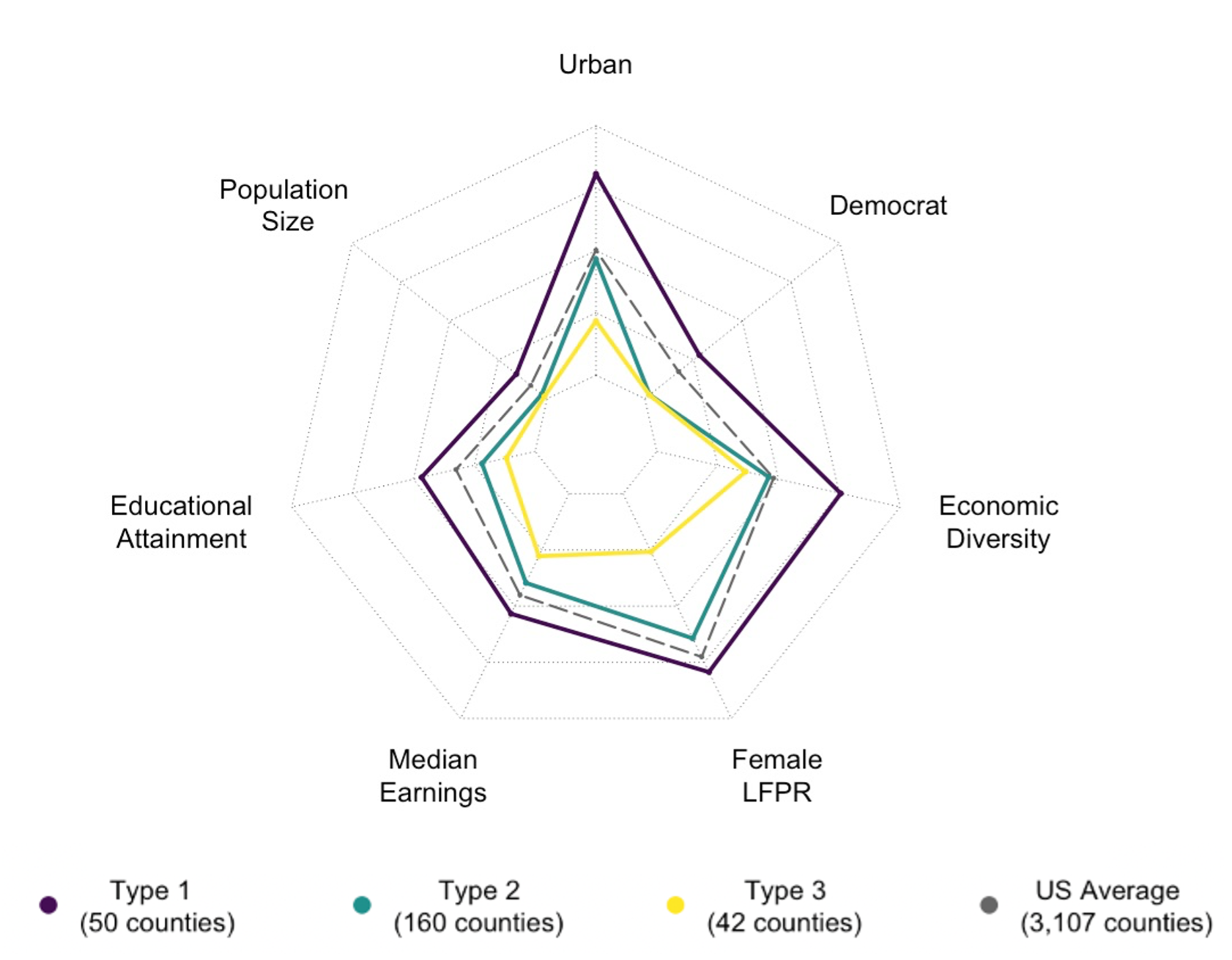}
\caption{\emph{The outer (inner) limit of the radar plot represents the maximum (minimum) value of each characteristic present in the dataset of coal counties. The variable for political affiliation (2016 and 2020 national election returns) is included in this radar plot for illustrative purposes. It is not included in the clustering method as nearly all counties with active coal mines between 2002 and 2019 voted for the Republican Party in the 2016 and 2020 elections. Average values for each indicator per group and the US overall used to generate this plot are reported in detail in Appendix D of the Supplementary Materials.}}
\label{fig:radar}
\end{figure}

\begin{figure}[h]
\centering
\includegraphics[width=15cm]{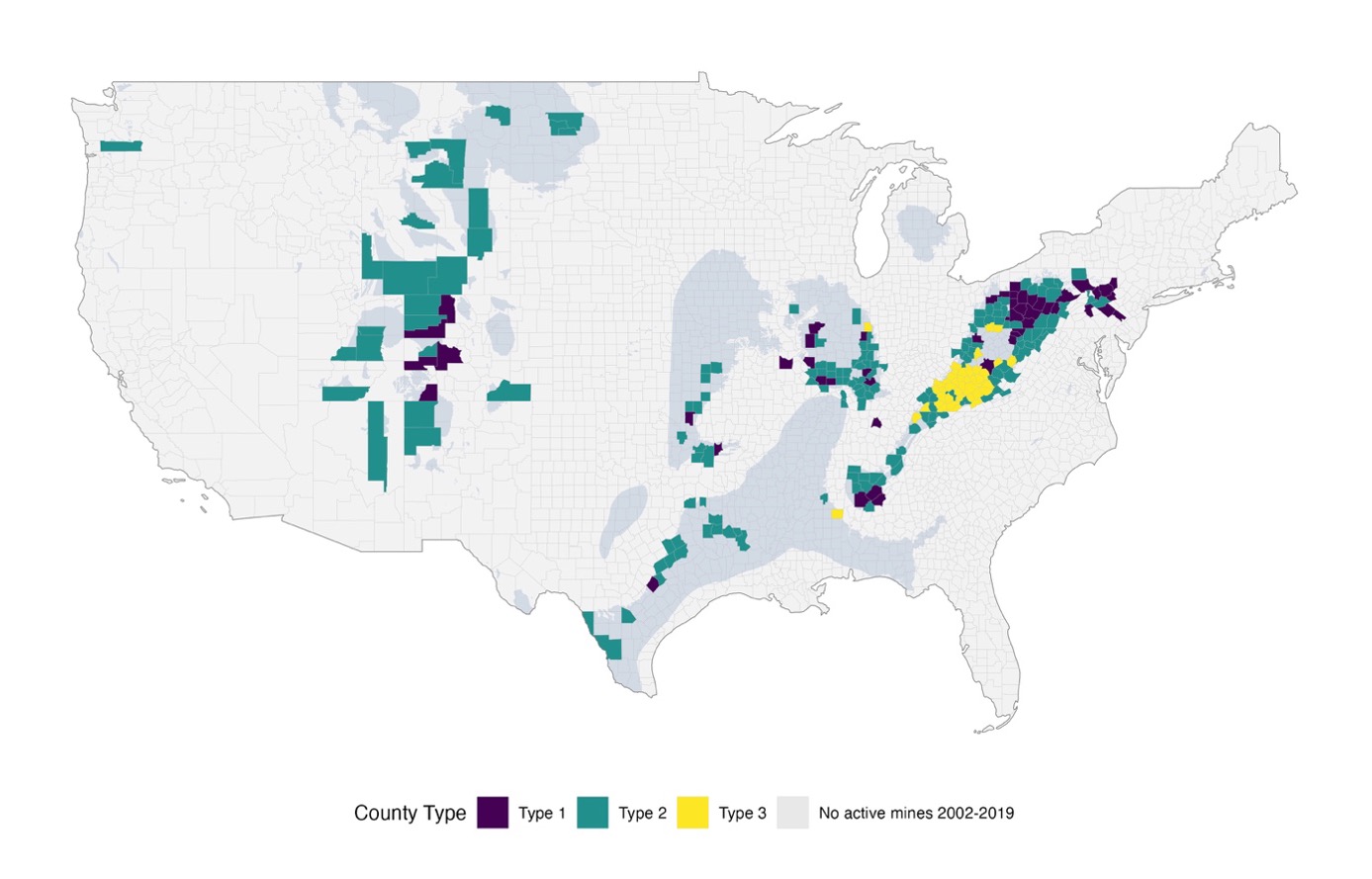}
\caption{\emph{Counties with active coal mines in the period 2002-2019 are color coded according to their “type” as defined by the agglomerative hierarchical clustering performed. Type 1 counties are considered “least vulnerable” due to their relatively stronger performance across the indicators (ie. urban with higher levels of income, educational attainment, economic diversity, female labor force participation, and population), selected in the construction of the typology whereas Type 3 counties are “most vulnerable.” The shaded blue areas represent, from left to right, the Western, Interior, and Appalachian coal basins generated using data available from the US Geological Survey.}}
\label{fig:typology-map}
\end{figure}

Figure \ref{fig:radar} provides a visual comparison of each county type identified through an application of hierarchical agglomerative clustering on the described set of county characteristics. Figure \ref{fig:radar} presents the mean values of each indicator per county type and for the US overall. These mean characteristics are reported in Appendix D of the Supplementary Materials. The geographical distribution of these types is represented in Figure \ref{fig:typology-map}.
\par

The three county “types” fall into a spectrum ranging from large and urban with high levels of educational attainment, income, female labor force participation, and economic diversity (Type 1) to small and rural with lower levels of the subsequent indicators (Type 3). This pattern is perhaps unsurprising to most readers as the economic and social indicators outlined tend to decrease between urban and rural counties. However, one of the more significant results is that Type 3, and to a lesser extent Type 2, coal counties fall well below the national average for all observed indicators that are likely to aid in an eventual transition. Furthermore, Figure \ref{fig:typology-map} demonstrates that Type 3 counties (yellow) are concentrated in the Appalachian coal region, indicating a high risk of regional decline which is already being observed in the area. The fact that Type 3 counties are clustered in one region and share most of their borders with each other or Type 2 counties (apart from one Type 1 county) also indicates a geographical limitation to accessible recovery or transition options should there be a need for laid-off workers migrate. 

\subsection{Putting the Typology to Work}

\begin{figure}[h]
\centering
\includegraphics[width=15cm]{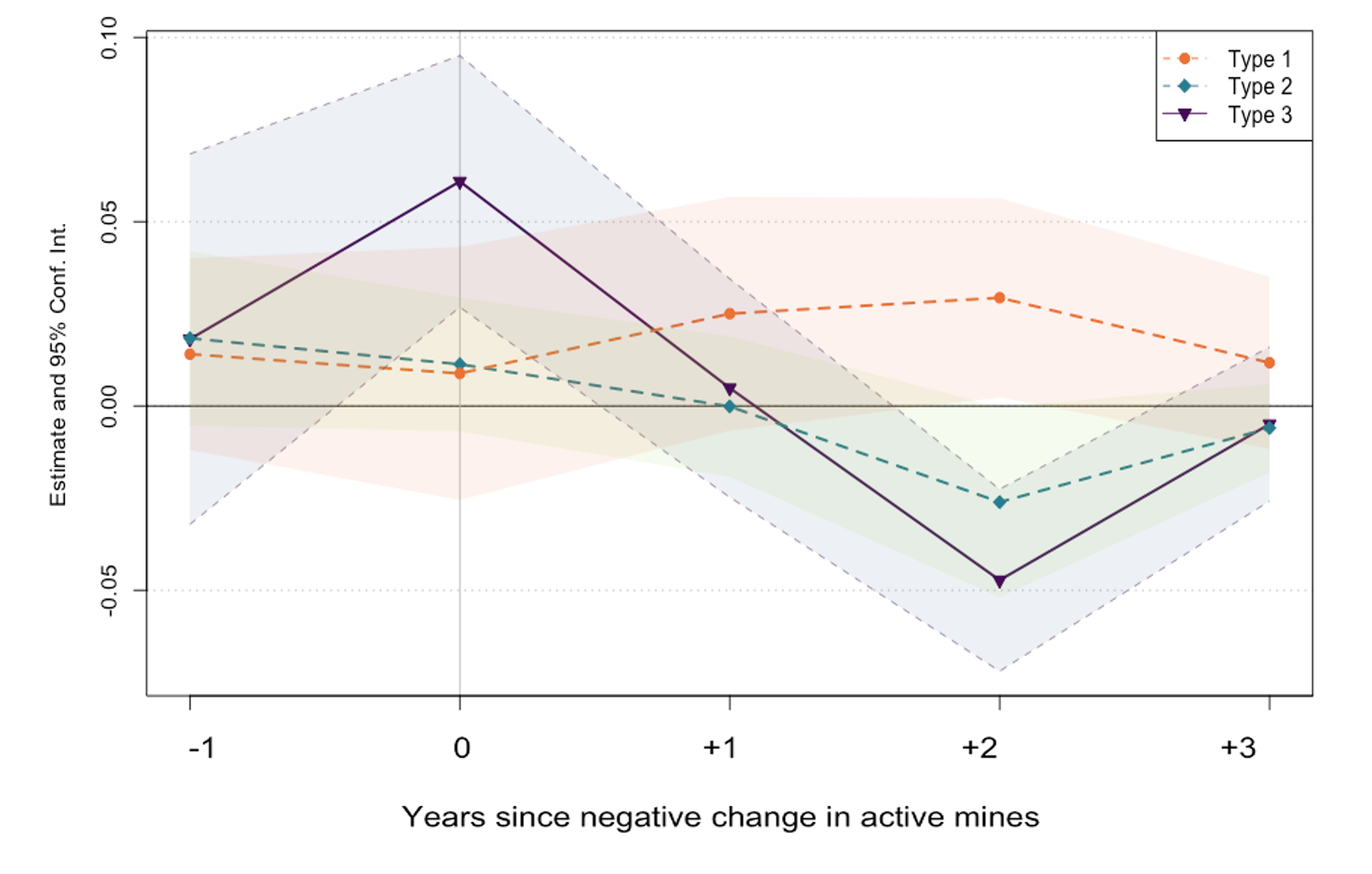}
\caption{\emph{As in Figure \ref{fig:coef_standard}, Figure \ref{fig:gfe} presents the change in unemployment rate and associated 95\% confidence interval resulting from a change in active mines. The sign of the coefficient estimate has been flipped to demonstrate the impact of a negative change in active mines. Each plotted line and confidence interval represents the response to a change in active mines by county type, as defined in Appendix D.2 in Supplementary Materials.}}
\label{fig:gfe}
\end{figure}

To put this typology to work, Model 1 was re-estimated using grouped fixed effects, allowing heterogeneous treatment effects for each county type. This enables us to determine whether the scale and duration of unemployment rate responses to changes in active coal mines differ across county types. Figure \ref{fig:gfe} demonstrates the coefficient estimates across our selected time horizon. Although Model 1 only includes the independent variables listed in Model 1 of Table \ref{tbl:models-specs}, the additional lags and leads displayed in Figure \ref{fig:gfe} do not greatly impact the estimates at times t = t, t+1, and t+2 and are therefore included in the visual below for additional information. Regression results from this grouped fixed effects application of Model 1 and expanded one-lead and four-lag models are reported in Appendix D.2 of the Supplementary Materials. 
\par

Most notably, Figure \ref{fig:gfe} illustrates that the greatest unemployment rate increases are observed in Type 3 counties, at a magnitude of 0.060 percentage points (1\% significance level) in response to a contemporaneous one-unit decrease in active mines. The coefficient estimates associated with Type 2 and 3 counties are all much smaller in magnitude and not statistically significant. 

\section{Discussion}
In this study, we examine employment shocks in US coal-mining counties undergoing decarbonization-related transitions. We corroborate our results using various panel econometric methods designed to address unobserved heterogeneity and cross-sectional dependence, including TWFE OLS, spatial effects, heterogeneous trends, and grouped fixed effects estimators. We find that a one-unit decline in active mines increases county-level unemployment by between 0.056 to 0.064 percentage points, with little change a year later and a small ‘recovery’ two years later. We further determine that this ‘recovery’ is likely not attributable to a recovery in employment. Given the irreducible element of unpredictability in county unemployment rates, especially in relation to the closure of an often proportionally small sector like coal, it is highly significant that our results show consistent estimates of regression coefficients across all models and reinforces our confidence in using these estimates to guide further reflections on policy responses.  
\par
Spatial and heterogeneous trend models show how the TWFE OLS approach underestimates unemployment effects. Incorporating spatial diffusion shows the risk of regional spill-over. Incorporating an asymmetric treatment estimation shows that opening new mines has no obvious influence on unemployment, but mine closures do. This result questions the belief that maintaining or boosting the coal sector will contribute to employment and reveals that mine closures have a stronger impact on unemployment than was initially detected in Model 1. This latter result provides crucial information for improving proposals - such as the one recently proposed by Senator Joe Manchin – to reform federal and state permit approval processes in the mining and energy sectors \cite{holzman2022a}. In particular, efforts to speed up and streamline permit approval process, as advocated in Manchin’s proposal, would very likely boost green transition mining while also artificially extending the coal sector’s operations beyond what its market outlook and the exigencies of climate change could conceivably permit. As the country awaits the congressional verdict on how the government will balance the trade-offs between issuing fast and efficient energy permitting on the one hand and appropriate environmental impact accounting on the other, future research could investigate how it may be possible to ramp up 'green' mining of minerals such as copper and silicon – which will be necessary to acquire the raw inputs enabling domestic manufacture of green technologies –  while at the same time avoiding continued reliance on coal energy and redeploying, where possible, laid-off coal miners interested in supporting the nation’s next phase of green industrialization \cite{galbraith2020a}.
\par
We further evaluated the influence of renewable energy investments on job losses to evaluate claims that green sectoral investments should be the principal means of aiding transitioning coal mining regions.  Unfortunately, we were unable to detect significant effects of these investments on county unemployment rates. Determining whether this is due to the lack of a systematic relationship between the two variables or due simply to a lack of sufficiently large investment volumes during the period examined is outside the scope of this study. Nevertheless, we propose ramping up such investments, not only to decarbonize the US economy, but also to enhance the potentialities of new employment in coal counties with already low economic diversification.  In certain localities, such initiatives have been beneficial, and the Inflation Reduction Act is encouraging on that front, even though it ignores the likely utility of other complimentary policy measures \cite{lakhani2022a}.
\par
Lastly, the typology offered in this paper reveals that most coal counties are slated to experience outsized socio-economic struggles following large-scale decarbonization compared to the nation's average performance across selected parameters. Empirical research shows that Type 3 counties, mostly in Appalachia, are the most heavily impacted.  Most transitional assistance and social reconstruction efforts should focus on these counties.
\par
These findings provide information crucial for making policy prescriptions since they enable better consideration of the specific needs of communities most impacted. First, low educational attainment and female labour force participation in Type 2 and 3 counties underscore the need for reskilling, retraining, and subsidized community college or vocational training for women. Subsidized day-care and after-school programs will be required so parents may attend these trainings after work. Second, the low level of economic diversity in Type 2 and 3 counties may hinder investment in other sectors. Investments that counteract Type 2 and 3 counties' poor economic diversity therefore ought to be considered alongside active labour market policies. Public subsidies may be needed to stimulate investment in low-diversity areas. The IRA's proposed reinvestment in energy infrastructure looks wise in this respect. Future research can investigate which policy instruments would most effectively mitigate against avoidable socio-economic and psychological hardship in places "left behind" due to bad investment conditions and low economic diversity. 
\par
Interestingly, even in Type 3 counties, median incomes are near the national average, so direct income support may not be the greatest use of public or private funds. However, income support to cover health care costs could be necessary during lapses in insurance, a claim demanded by the United Mine Workers of America (UMWA) \cite{umwa2017a}.
\par
Our econometric analysis indicates that the greatest policy difficulty may not be a time-persistent unemployment shock but rather its spatial ripple effect. The rural-urban split between Type 2 and 3 counties and Type 1 counties would likely require substantial inter-regional mobility or transportation upgrades to help laid-off workers find new jobs. Due to the preponderance of Type 3 counties, Appalachia will likely have the highest investment costs. In severe cases, relocation support should be explored for unemployed workers in rural areas with minimal economic diversity and wherever inter-regional transit is extremely expensive or impossible.
\par
A main risk of a poorly managed transition is the potential to erode support for future environmental action. Recent research on popular acceptance of environmental measures in the US indicates that environmental policy bundled with social safeguards might boost public support, offering further grounds for the social and political practicality of a Just Transition agenda \cite{piggot2019a, bergquist2020a}. Incorporating Just Transition concepts into decarbonization policies may boost support for environmental programmes generally. The UMWA seems to reflect this sentiment, as it has expressed a willingness to embrace a green transition provided that affected workers are supported along the way \cite{daly2021a}.
\par
Finally, to the disservice of US communities still struggling with the effects of coal’s decline, the 'situation' of US coal employees is increasingly being used as an example of failed Just Transition aspirations, with academic researchers and public commentators asking how the transition from oil and gas might avoid similar issues. This study reiterates that coal communities deserve sustained attention while also outlining a framework of inquiry that could inform future research focusing on unemployment shocks following oil and gas sector decarbonization.
\section{Conclusion}
The research presented herein is crucial because it elucidates the immediate effects of mine closures on coal communities and demonstrates the urgent requirement for transition support following an unemployment shock caused by decarbonization. Coal-dependent counties are typically rural areas with limited economic opportunities, low rates of female labour force participation, and low levels of educational attainment. Policies that aim to counteract the transitional risks, vulnerabilities, and hardships experienced in these disproportionately affected communities can achieve maximum effectiveness provided they are strategically targeted and sufficiently farsighted to minimise the geographic spill-over of negative repercussions. This research has paved the way for the development of empirically grounded, context-dependent ameliorative strategies. This research has also shed light on how the provisions of the recently enacted Inflation Reduction Act might benefit communities that rely on fossil fuels, showing that while smarter green economy investment targeting could help these communities weather the inevitable green transition, broader and more concerted policy provisions are needed to ensure a Just Transition.

\bibliographystyle{abbrvurl}
\bibliography{wp_jtcoal}

\section{Supplementary Materials}
\singlespacing

\renewcommand{\thetable}{S\arabic{table}}
\renewcommand{\thefigure}{S\arabic{figure}}
\date{}

\begin{appendices}

\section{Summary Statistics and Data Sources}
\subsection{Econometric Models}

\begin{table}[H] 
\centering 
  \caption{\large Indicators and Data Sources} 
    \begin{tabular}{|l|l|l|}
    \hline
        \textbf{Indicator} & \textbf{Data Source} & \textbf{Survey/Dataset} \\ \hline
        \addlinespace
        Population & US Department of Commerce  & Local Area Personal Income  \\ 
        & Bureau of Economic Analysis (BEA)& and Employment Statistics \\\hline
        \addlinespace
        Labour Force & ~ & Local Area \\ 
        Employed Persons &US Bureau of Labor Statistics (BLS)&Unemployment Statistics \\ 
        Unemployed Persons &~&~\\
        Unemployment Rate &~&~\\ \hline
        \addlinespace
        Active mines & US Energy Information Administration (EIA) & Mine Employment and   \\ 
        &and the US Mine Safety and Health & Coal Production Data\\
        &Administration (MSHA)&~\\\hline
        \addlinespace
        Renewable Energy & US Department of Agriculture & Energy Investment Report \\
        Investments & ~& ~\\\hline
        \addlinespace
        County Adjacency Matrix & US Census Bureau & County Adjacency File \\
        \hline
    \end{tabular}
\end{table}
\begin{table}[H] 
\centering 
  \caption{\large Summary Statistics: Contiguous US Counties} 
\begin{tabular}{@{\extracolsep{5pt}}lccccccc} 
\\[-1.8ex]\hline 
\hline \\[-1.8ex] 
Statistic & \multicolumn{1}{c}{N} & \multicolumn{1}{c}{Mean} & \multicolumn{1}{c}{St. Dev.} & \multicolumn{1}{c}{Min} & \multicolumn{1}{c}{Pctl(25)} & \multicolumn{1}{c}{Pctl(75)} & \multicolumn{1}{c}{Max} \\ 
\hline \\[-1.8ex] 
Active Mines & 55,296 & 0.462 & 3.723 & 0 & 0 & 0 & 114 \\ 
Unemployment Rate & 55,296 & 6.130 & 2.720 & 0.800 & 4.200 & 7.500 & 29.400 \\ 
Employed Persons & 55,296 & 46,504 & 149,661 & 34 & 4,799 & 30,140 & 4,888,581 \\ 
Unemployed Persons & 55,296 & 2,993 & 11,183 & 3 & 293 & 1,988 & 621,950 \\ 
Labour Force & 55,296 & 49,498 & 159,938 & 38 & 5,128 & 32,055 & 5,122,843 \\ 
Population & 55,296 & 99,440 & 318,828 & 55 & 11,223 & 67,553 & 10,105,708 \\ 
RE Investments (USD)* & 55,296 & 114,142 & 2,670,894 & 0 & 0 & 0 & 250,000,000 \\ 
Real GDP & 55,296 & 5,140,646 & 22,046,739 & 7,648 & 342,473 & 2,550,861 & 726,943,301 \\ 
Real GDP Per Capita & 55,296 & 50.84 & 463.90 & 5.79 & 25.90 & 46.76 & 59,848.92 \\ 
Rural-Urban Code & 55,296 & 5.1 & 2.7 & 1 & 3 & 7 & 9 \\ 
Rural-Urban (binary) & 55,296 & 0.65 & 0.48 & 0 & 0 & 1 & 1 \\ 
RE Inv. (prop. of Real GDP)** & 55,296 & 0.091 & 3.073 & 0 & 0 & 0 & 504 \\ 
\hline \\[-1.8ex] 
\end{tabular} 
\end{table}
\emph{*RE Investments (USD)}: Level of renewable energy investment in US dollars. \\
\emph{*RE Inv. (prop. of Real GDP)}: Level of renewable energy investment as proportion of county real GDP. \\

\begin{table}[H] 
\centering 
  \caption{\large Summary Statistics: US Coal Counties}  
\begin{tabular}{@{\extracolsep{5pt}}lccccccc} 
\\[-1.8ex]\hline 
\hline \\[-1.8ex] 
Statistic & \multicolumn{1}{c}{N} & \multicolumn{1}{c}{Mean} & \multicolumn{1}{c}{St. Dev.} & \multicolumn{1}{c}{Min} & \multicolumn{1}{c}{Pctl(25)} & \multicolumn{1}{c}{Pctl(75)} & \multicolumn{1}{c}{Max} \\ 
\hline \\[-1.8ex] 
Active Mines & 4,518 & 5.653 & 11.847 & 0 & 1 & 5 & 114 \\ 
Unemployment Rate & 4,518 & 6.946 & 2.520 & 2 & 5.2 & 8.3 & 21 \\ 
Employed Persons & 4,518 & 29,070 & 58,839 & 816 & 6,859 & 27,426 & 622,714 \\ 
Unemployed Persons & 4,518 & 1,933 & 3,535 & 32 & 477 & 1,890 & 46,564 \\ 
Labour Force & 4,518 & 31,003 & 62,187 & 870 & 7,392 & 29,621 & 651,926 \\ 
Population & 4,518 & 64,867 & 119,766 & 1,836 & 16,721 & 64,537 & 1,265,577 \\ 
RE Investments (USD)* & 4,518 & 27,056 & 333,440 & 0 & 0 & 0 & 10,005,017 \\ 
Coal Production (short tons) & 4,518 & 4,057,556 & 22,253,668 & 0 & 0 & 3,564,906 & 415,924,096 \\ 
Real GDP & 4,518 & 3,045,039 & 7,913,160 & 47,597 & 590,750 & 2,623,148 & 92,984,370 \\ 
Real GDP Per Capita & 4,518 & 41.688 & 25.763 & 8.221 & 26.348 & 48.551 & 244.161 \\ 
Rural-Urban Code & 4,518 & 5.116 & 2.376 & 1 & 3 & 7 & 9 \\ 
Rural-Urban (binary) & 4,518 & 0.697 & 0.460 & 0 & 0 & 1 & 1 \\ 
TAA Allocation (USD) & 4,518 & 14,597,789 & 20,533,595 & 0 & 0 & 23,427,230 & 117,476,517 \\ 
RE Inv. (prop. of Real GDP)** & 4,518 & 0.016 & 0.241 & 0 & 0 & 0 & 9 \\ 
\hline \\[-1.8ex] 
\end{tabular} 
\end{table} 

\begin{table}[H] 
\centering 
  \caption{\large Summary Statistics of Transformed Variables: Contiguous US Counties} 
\begin{tabular}{@{\extracolsep{5pt}}lccccccc} 
\\[-1.8ex]\hline 
\hline \\[-1.8ex] 
Statistic & \multicolumn{1}{c}{N} & \multicolumn{1}{c}{Mean} & \multicolumn{1}{c}{St. Dev.} & \multicolumn{1}{c}{Min} & \multicolumn{1}{c}{Pctl(25)} & \multicolumn{1}{c}{Pctl(75)} & \multicolumn{1}{c}{Max} \\ 
\hline \\[-1.8ex] 
$\Delta$ Unemployment Rate & 55,296 & $-$0.061 & 1.228 & $-$8.600 & $-$0.700 & 0.300 & 13.500 \\ 
$\Delta$ Active Mines\textsubscript{t} & 55,296 & $-$0.015 & 0.605 & $-$30 & 0 & 0 & 20 \\ 
$\Delta$ (log) Real GDP & 55,295 & 0.017 & 0.088 & $-$1.122 & $-$0.017 & 0.048 & 1.386 \\ 
$\Delta$ (log) Real GDP per capita & 55,295 & 0.014 & 0.088 & $-$1.126 & $-$0.020 & 0.043 & 1.391 \\ 
$\Delta$ (log) Employed Persons & 55,296 & 0.001 & 0.037 & $-$0.596 & $-$0.013 & 0.018 & 1.022 \\ 
$\Delta$ (log) Unemployed Persons & 55,296 & $-$0.013 & 0.177 & $-$0.852 & $-$0.124 & 0.057 & 1.305 \\ 
$\Delta$ (log) Labour Force & 55,296 & 0.001 & 0.034 & $-$0.562 & $-$0.013 & 0.015 & 0.986 \\ 
$\Delta$ {(log) Population} & 55,295 & 0.003 & 0.014 & $-$0.425 & $-$0.005 & 0.009 & 0.290 \\ 
REE $\geq 0.1$\% of GDP & 55,296 & 0.127 & 0.333 & 0 & 0 & 0 & 1 \\ 
\hline \\[-1.8ex] 
\end{tabular} 
\end{table}

\begin{table}[H] 
\centering 
  \caption{\large Summary Statistics of Transformed Variables: US Coal Counties Subset} 
\begin{tabular}{@{\extracolsep{5pt}}lccccccc} 
\\[-1.8ex]\hline 
\hline \\[-1.8ex] 
Statistic & \multicolumn{1}{c}{N} & \multicolumn{1}{c}{Mean} & \multicolumn{1}{c}{St. Dev.} & \multicolumn{1}{c}{Min} & \multicolumn{1}{c}{Pctl(25)} & \multicolumn{1}{c}{Pctl(75)} & \multicolumn{1}{c}{Max} \\ 
\hline \\[-1.8ex] 
$\Delta$ Unemployment Rate & 4,518 & $-$0.056 & 1.322 & $-$4.600 & $-$0.793 & 0.400 & 8.300 \\ 
$\Delta$ Active Mines\textsubscript{t} & 4,518 & $-$0.182 & 2.109 & $-$30 & 0 & 0 & 20 \\ 
$\Delta$ (log) Real GDP & 4,518 & 0.010 & 0.081 & $-$0.604 & $-$0.022 & 0.038 & 1.342 \\ 
$\Delta$ (log) Real GDP per capita & 4,518 & 0.010 & 0.080 & $-$0.594 & $-$0.021 & 0.037 & 1.351 \\ 
$\Delta$ (log) Employed Persons & 4,518 & $-$0.002 & 0.035 & $-$0.440 & $-$0.015 & 0.014 & 0.214 \\ 
$\Delta$ (log) Unemployed Persons & 4,518 & $-$0.014 & 0.179 & $-$0.479 & $-$0.120 & 0.051 & 1.077 \\ 
$\Delta$ (log) Labour Force & 4,518 & $-$0.003 & 0.031 & $-$0.396 & $-$0.015 & 0.012 & 0.185 \\ 
$\Delta$ (log) Population & 4,518 & $-$0.0004 & 0.010 & $-$0.098 & $-$0.006 & 0.004 & 0.066 \\ 
REE $\geq 0.1$\%  of GDP & 4,518 & 0.097 & 0.296 & 0 & 0 & 0 & 1 \\ 
\hline \\[-1.8ex] 
\end{tabular} 
\end{table} 

\subsection{Typology}

\begin{table}[H] 
\centering 
  \caption{\large Indicators and Data Sources} 
    \begin{tabular}{|l|l|l|l|}
    \hline
        \textbf{Characteristic}&\textbf{Indicator} & \textbf{Data Source} & \textbf{Survey/Dataset} \\\hline
        \addlinespace
        Rural vs. Urban & Rural-Urban Codes& US Department of  & Rural-Urban Continuum \\
        &&Agriculture's Economic &Codes 2013 \\
        &&Research Service& \\\hline
        \addlinespace
        Population Size & 2019 Population Estimate & US Census Bureau& Population and Housing \\
        & & & Unit Estimates\\\hline
        \addlinespace
        Educational& Percentage of population  &US Census Bureau& American Community Survey \\
        Attainment& aged 25-64 with at least & & \\
        &  a bachelor’s degree& & \\\hline
        \addlinespace
        Economic Security& Median earnings (USD)&US Census Bureau& American Community Survey\\
        & & & \\\hline
          \addlinespace
        Female Labor &Labor force participation rate &US Census Bureau& American Community Survey\\
        Force Participation& of the female population & & \\
        Rate& aged 25-64 years& & \\\hline
          \addlinespace
        Economic Diversity&Economic Diversity Index&Chmura Economics &Economic Diversity Index \\
        & & and Analytics& \\\hline
          \addlinespace
        Political Attitudes&2016 and 2020  &MIT Election Data  &County Presidential \\
        &Election Returns&and Science Lab&Election Returns 2000-2020\\ \hline
    \end{tabular}
\end{table}

\begin{center}
\large Typology: Motivation for Indicator Choices
\end{center}

\textbf{Summary}: County demographic, social, economic, and political characteristics that have previously been identified as having potential to affect or determine an area’s ability or potential to transition smoothly during and after and economic shift or shock were collected.

\textbf{Rural vs. Urban}: First, whether a county is more urban or rural can affect the proximity of newly unemployed workers formerly in brown industries to other job and education opportunities but will also impact identity construction surrounding extractive industries and limit accessibility to policymakers and other resources intended to support achieving a Just Transition.\\

\textbf{Population Size}: Second, population size is an important predictor of political reach and voice as well as economic activity and opportunities.\\

\textbf{Educational Attainment}: Third, the average level of educational attainment will impact qualifications and skill levels of individuals looking for new job opportunities, particularly in the face of a potential skills-biased structural decarbonization. This may be of particular concern in resource-rich regions, as job opportunities presented in extractive industries and natural resource production have been found to provide a disincentive to education, especially among young men.\\

\textbf{Economic Security}: Next, median earnings was used to proxy the level of economic security of a county. A lower level of economic security can imply great hardship for individuals and households facing periods of unemployment without savings to fall back on.\\

\textbf{Female Labor Force Participation Rate}: Next, the coal mining industry, much like the fossil fuel industry overall, has a largely male-dominated workforce meaning the prospect of lay-offs of a male-dominated workforce will leave many households reliant on the income or labor of females. Therefore, boosting female employment in other industries might provide crucial income support during periods of transition for the implicated partners. This effect is proxied in the typology by female labor force participation among women between 25 and 64 years of age.\\

\textbf{Economic Diversity}: Furthermore, economic diversity, or the mix of industries providing employment within a county, is an important determinant not only of the proximate available job opportunities for workers to transition into but also potential sites of investment to boost economic opportunities, whether green or otherwise.\\

\textbf{Political Attitudes}: Lastly, political attitudes or party affiliation will dictate the type of messaging surrounding transitions that will either corrode or boost public support and willingness to engage with implemented support measures. Furthermore, as ambitions relating to environmental regulation as well as investments in renewable energy and energy efficiency industries are generally advanced by the US Democratic party, political party affiliation will likely dictate the kinds of opportunities that will be created for fossil fuel workers facing unemployment risks. Originally, political attitudes were intended to be proxied by county returns in the 2016 and 2020 presidential elections. However, only 19 of 252 coal counties voted for the Democratic party in the 2016 and 2020 presidential elections and this was therefore excluded from the clustering analysis as Republican party affiliation seems to be a unifying characteristic among coal counties. \\

\begin{table}[H] 
\centering 
  \caption{\large Summary Statistics: Stylized Typology Characteristics}
\begin{tabular*}{\textwidth}{@{}  
                            @{\extracolsep{\fill}}lccccccc}
\\[-1.8ex]\hline 
\hline \\[-1.8ex] 
Statistic & \multicolumn{1}{c}{N*} & \multicolumn{1}{c}{Mean} & \multicolumn{1}{c}{St. Dev.} & \multicolumn{1}{c}{Min} & \multicolumn{1}{c}{Pctl(25)} & \multicolumn{1}{c}{Pctl(75)} & \multicolumn{1}{c}{Max} \\ 
\hline \\[-1.8ex] 
2013 Rural-Urban Code & 252 & 5.1 & 2.4 & 1 & 3 & 7 & 9 \\ 
~&~&~&~&~&~&~& \\
Population Size & 252 & 64,655.3 & 121,468.2 & 1,959 & 16,684.5 & 64,320.2 & 1,216,045 \\
~&~&~&~&~&~&~& \\
Educational Attainment & 252 & 18.2 & 7.7 & 5.4 & 13.6 & 20.7 & 56.2 \\ 
\-\hspace{.5cm}(\%; aged 25-64) &&&&&&& \\
Median Earnings (USD) & 252 & 34,204.2 & 4,935.0 & 20,268 & 31,357 & 36,705.2 & 54,754 \\
~&~&~&~&~&~&~& \\
Female Labour Force Participation & 252 & 64.6 & 8.9 & 37.9 & 60.3 & 71.1 & 81.7 \\ 
\-\hspace{.5cm}(\%; aged 25-64) &&&&&&& \\
Chmura Diversity Index & 252 & 1.0 & 0.2 & 0.5 & 0.9 & 1.1 & 1.5 \\ 
~&~&~&~&~&~&~& \\
Voted for the Republican& 252 & 0.9 & 0.3 & 0 & 1 & 1 & 1 \\
 Party in the 2020 Election &~&~&~&~&~&~& \\
Voted for the Republican & 252 & 0.9 & 0.3 & 0 & 1 & 1 & 1 \\
 Party in the 2016 Election &~&~&~&~&~&~& \\
\hline \\[-1.8ex] 
\end{tabular*} \\
\end{table} 

Note: The state of Virginia’s counties are measured differently across data sources. More precisely, the BEA combines independent cities of Virginia with adjacent counties thus reporting less county observations than in the remaining data sources. For the econometric applications, this discrepancy was overcome by adjusting the remaining data sources to match the county identification codes provided by the BEA. Conventional FIPS codes were used for the typology analysis to allow for the inclusion of the relevant indicators with minimal additional manipulation. Only one county retained in the typology analysis is implicated in the BEA-Census Bureau discrepancy, thus the impact of this choice on the results is deemed minimal. Therefore, in the subsequent analysis, econometric applications make use of 3,072 county observations versus the 3,107 counties observed in the typology dataset. Additionally, for the subset of coal counties, the number of counties included in the typology (252) differs from the number identified in the econometric analysis (251) because Wise County, Virginia and Norton City, Virginia  are considered separately by the various entities reporting data for the typology characteristics. They are combined into one county area by the Bureau of Economic Analysis whose method was adopted as the standard for the econometric analysis. \\

Alaska, Hawaii, and the District of Columbia (DC) were excluded from the analysis. Hawaii does not produce coal. Indicators for DC are inconsistently reported across the datasets involved in this analysis. During the 18-year time period analyzed, Alaska’s county equivalents were divided and re-grouped making it difficult to compare available data between time periods. The exclusion of DC is considered a limitation in this work.

\section{Econometric Models}
\subsection{Two-way fixed effects (TWFE) OLS models}
\subsubsection{Level indicators}
\begin{table}[H]
\centering
\caption{\large Impact of the number of active mines on county unemployment rate (level)\\
\scriptsize Using a TWFE OLS regression model with various combinations of time lags and leads}
\begin{tabular}{lcccccc}
   \tabularnewline\midrule\midrule
   Dependent Variable: & \multicolumn{6}{c}{Unemployment Rate}\\
   Model:              & (1)             & (2)            & (3)             & (4)             & (5)            & (6)\\
   ~ & \multicolumn{3}{c}{All US Counties} & \multicolumn{3}{c}{Coal Counties Only} \\
   \midrule \emph{Variables} &   &   &   &   &   &  \\
   Active Mines\textsubscript{t+3} &                 &                & -0.0284.   &                 &                & -0.0177\\
                       &                 &                & (0.0134)        &                 &                & (0.0116)\\
   Active Mines\textsubscript{t+2} &                 & -0.0293$^{*}$ & -0.0030         &                 & -0.0220.  & -0.0059\\
\end{tabular}
\end{table}

\begin{table}[H]
\centering
\caption{\large Impact of the number of active mines on county employment indicators (levels)\\
\scriptsize Using a TWFE OLS regression model with two time lags on all contiguous US counties}
\begin{tabular}{lcccc}
   \tabularnewline\midrule\midrule
   Dependent Variables: & Employed & Unemployed  & Labour   & Population\\
   ~ & Persons & Persons     & Force   & ~\\
   Model:               & (1)              & (2)            & (3)              & (4)\\
   \midrule \emph{Variables} &   &   &   &  \\
   Active Mines         & 12.69            & -1.205         & 11.48            & 206.3$^{*}$\\
                        & (31.64)          & (8.444)        & (25.46)          & (88.22)\\
   Active Mines (t-1)   & 5.954            & -10.81         & -4.855           & 3.627\\
\end{tabular}
\end{table}

\begin{table}[H]
\centering
\caption{\large Impact of the number of active mines on county employment indicators (log levels excl. UER)\\
\scriptsize Using a TWFE OLS regression model with two time lags on all contiguous US counties}
\begin{tabular}{lccccc}
   \tabularnewline\midrule\midrule
   Dependent Variables: & Unemployment & Employed & Unemployed & Labour   & Population\\
~ & Rate & Persons (log) & Persons (log)      & Force (log)       & (log)\\
   Model:               & (1)               & (2)            & (3)                    & (4)            & (5)\\
   \midrule \emph{Variables} &   &   &   &   &  \\
   Active Mines         & -0.0955$^{***}$   & 0.0015$^{*}$  & -0.0103$^{***}$        & 0.0004         & 0.0012\\
                        & (0.0205)          & (0.0006)       & (0.0020)               & (0.0006)       & (0.0007)\\
   Active Mines\textsubscript{t-1}   & -0.0035           & 0.0012.   & $-9.24\times 10^{-5}$ & 0.0012         & 0.0003\\
\end{tabular}
\end{table}

\begin{table}[H]
\centering
\caption{\large Impact of a change in active mines on county employment indicators (log levels excl. unemployment rate)\\
\scriptsize Using a TWFE OLS regression model on all contiguous US counties}
\begin{tabular}{lccccc}
   \tabularnewline\midrule\midrule
   Dependent Variables:   &Unemployment   & Employed & Unemployed & Labour       & Population\\
~ & Rate  & Persons (log) & Persons (log) & Force (log)      & (log)\\
   Model:                    & (1)               & (2)            & (3)              & (4)            & (5)\\
   \midrule \emph{Variables} &   &   &   &   &  \\
   $\Delta$ Active Mines\textsubscript{t}       & -0.0835$^{***}$   & 0.0007         & -0.0081$^{***}$  & -0.0002        & 0.0003\\
                             & (0.0198)          & (0.0006)       & (0.0020)         & (0.0007)       & (0.0008)\\
   $\Delta$ Active Mines\textsubscript{t-1} & -0.0884$^{***}$   & 0.0020$^{**}$ & -0.0082$^{***}$  & 0.0010.   & 0.0006\\
\end{tabular}
\end{table}

\subsubsection{First-difference indicators}

\begin{table}[H]
\centering
\caption{\large Impact of change in active mines on the change in unemployment rate\\
\scriptsize Using a TWFE OLS regression model on all contiguous counties and coal counties with various combinations of time lags and leads}
\begin{tabular}{lcccccc}
   \tabularnewline\midrule\midrule
   Dependent Variable: & \multicolumn{6}{c}{$\Delta$ Unemployment Rate}\\
   Model:                      & (1)             & (2)             & (3)             & (4)             & (5)             & (6)\\
   ~ & \multicolumn{3}{c}{All US Counties} & \multicolumn{3}{c}{Coal Counties Only} \\
   \midrule \emph{Variables} &   &   &   &   &   &  \\
   $\Delta$ Active Mines\textsubscript{t+3}           &                 &                 & -0.0186$^{*}$  &                 &                 & -0.0087\\
                               &                 &                 & (0.0082)        &                 &                 & (0.0080)\\
   $\Delta$ Active Mines\textsubscript{t+2}           &                 & -0.0050         & -0.0056         &                 & -0.0068         & -0.0065\\
\end{tabular}
\end{table}

\begin{table}[H]
\centering
\caption{\large Impact of change in active mines on the change in county employment indicators (log excl. UER)\\
\scriptsize Using a TWFE OLS regression model on all contiguous US counties}
\begin{tabular}{lccccc}
   \tabularnewline\midrule\midrule
   Dependent Variables:                & $\Delta$ Unemployment & $\Delta$ Employed & $\Delta$ Unemployed & $\Delta$ Labour & $\Delta$ Population\\
~   &  Rate & Persons (log) & Persons (log) & Force (log) & (log)\\
   Model:                      & (1)                      & (2)                              & (3)                                & (4)                          & (5)\\
   \midrule \emph{Variables} &   &   &   &   &  \\
   $\Delta$ Active Mines\textsubscript{t}         & -0.0563$^{***}$          & 0.0011$^{*}$                    & -0.0055$^{**}$                    & 0.0005                       & $-4.5\times 10^{-5}$\\
                               & (0.0141)                 & (0.0004)                         & (0.0016)                           & (0.0003)                     & ($7.38\times 10^{-5}$)\\
   $\Delta$ Active Mines\textsubscript{t-1}   & -0.0070                  & 0.0016$^{***}$                   & -0.0003                            & 0.0016$^{***}$               & 0.0001\\
\end{tabular}
\end{table}

\begin{table}[H]
\centering
\caption{\large Impact of change in active mines on changes in county employment indicators (log excl. UER)\\
\scriptsize Using a TWFE OLS regression model on US coal counties}
\begin{tabular}{lccccc}
   \tabularnewline\midrule\midrule
   Dependent Variables:                & $\Delta$ Unemployment & $\Delta$ Employed & $\Delta$ Unemployed & $\Delta$ Labour & $\Delta$ Population\\
~& Rate & Persons (log) & Persons (log) & Force (log)& (log)\\
   Model:                      & (1)                      & (2)                              & (3)                                & (4)                          & (5)\\
   \midrule \emph{Variables} &   &   &   &   &  \\
   $\Delta$ Active Mines\textsubscript{t}         & -0.0413$^{**}$          & 0.0008.                     & -0.0037$^{*}$                     & 0.0003                       & $-4.21\times 10^{-5}$\\
                               & (0.0124)                 & (0.0004)                         & (0.0013)                           & (0.0003)                     & ($5.96\times 10^{-5}$)\\
   $\Delta$ Active Mines\textsubscript{t-1}   & -0.0042                  & 0.0013$^{***}$                   & 0.0002                             & 0.0013$^{***}$               & $5.8\times 10^{-5}$\\
\end{tabular}
\end{table}

\begin{table}[H]
\centering
\caption{\large Impact of change in active mines in various time periods on the change in unemployment rate\\
\scriptsize Using a TWFE OLS regression model on all contiguous counties and coal counties using single time periods}
\begin{tabular}{lcccccccc}
   \tabularnewline\midrule\midrule
   Dependent Variable: & \multicolumn{8}{c}{$\Delta$ Unemployment Rate}\\
   Model:                      & (1)             & (2)             & (3)             & (4)             & (5)            & (6)            & (7)            & (8)\\
   \midrule \emph{Variables} &   &   &   &   &   &   &   &  \\
   $\Delta$ Active Mines\textsubscript{t}         & -0.0560$^{**}$  &                 &                 &                 & -0.0416$^{**}$ &                &                &   \\
                               & (0.0151)        &                 &                 &                 & (0.0127)       &                &                &   \\
   $\Delta$ Active Mines\textsubscript{t-1}   &                 & -0.0050         &                 &                 &                & -0.0029        &                &   \\
                               &                 & (0.0137)        &                 &                 &                & (0.0096)       &                &   \\
\end{tabular}
\end{table}

\begin{table}[H]
\centering
\caption{\large Impact of change in active mines on change in unemployment rate\\
\scriptsize Using a TWFE OLS regression model and binary interaction term for whether a county is rural or not (Rural = 1; Non-rural = 0)}
\begin{tabular}{lcc}
   \tabularnewline\midrule\midrule
   Dependent Variable: & \multicolumn{2}{c}{$\Delta$ Unemployment Rate}\\
   Model:                                    & (1)             & (2)\\
   \midrule \emph{Variables} &   &  \\
   $\Delta$ Active Mines\textsubscript{t}                       & -0.0308$^{*}$  & -0.0119\\
                                             & (0.0135)        & (0.0134)\\
   $\Delta$ Active Mines\textsubscript{t-1}                & -0.0371$^{*}$  & -0.0306.\\
                                             & (0.0163)        & (0.0160)\\
\end{tabular}
\end{table}

\begin{table}[H]
\centering 
  \caption{\large Asymmetric treatment on all US counties}
  \scriptsize{Performed using a TWFE OLS regression on all contiguous US counties.}
\begin{tabular}{lccc}
   \tabularnewline\midrule\midrule
   Dependent Variable: & \multicolumn{3}{c}{$\Delta$ Unemployment Rate}\\
   Model:                                             & (1)             & (2)             & (3)\\
   \midrule \emph{Variables} &   &   &  \\
   $\Delta$ Active Mines\textsubscript{t}                                & -0.0193         & -0.0752$^{**}$  &   \\
                                                      & (0.0201)        & (0.0235)        &   \\
   $\Delta$ Active Mines\textsubscript{t-1}                         & 0.0221          & -0.0301         &   \\
                                                      & (0.0285)        & (0.0219)        &   \\
\end{tabular}
\end{table}

\begin{table}[H]
\centering 
  \caption{\large Asymmetric treatment on subset of US coal counties} 
  \scriptsize{Performed using a TWFE OLS regression on US coal counties.}
\begin{tabular}{lccc}
   \tabularnewline\midrule\midrule
   Dependent Variable: & \multicolumn{3}{c}{Change Unemployment Rate}\\
   Model:                                             & (1)             & (2)             & (3)\\
   \midrule \emph{Variables} &   &   &  \\
   $\Delta$ Active Mines\textsubscript{t}                                & -0.0193         & -0.0752$^{**}$  &   \\
                                                      & (0.0201)        & (0.0235)        &   \\
   $\Delta$Active Mines\textsubscript{t-1}                          & 0.0221          & -0.0301         &   \\
                                                      & (0.0285)        & (0.0219)        &   \\
\end{tabular}
\end{table}

\begin{table}[H]
\centering 
  \caption{\large Regression Results: Impact of changes in active mines on unemployment rate accounting for county-level renewable energy investment}  
\begin{tabular}{lcc}
   \tabularnewline\midrule\midrule
   Dependent Variable: & \multicolumn{2}{c}{$\Delta$ Unemployment Rate}\\
   Model:                                                     & (1)            & (2)\\
   \midrule \emph{Variables} &   &  \\
   $\Delta$ Active Mines\textsubscript{t}                                        & -0.0640$^{**}$ & -0.0470$^{**}$\\
                                                              & (0.0163)       & (0.0144)\\
   $\Delta$ Active Mines\textsubscript{t-1}                                  & -0.0015        & -0.0001\\
                                                              & (0.0160)       & (0.0113)\\
\end{tabular}
\end{table}

\begin{table}[H]
\centering 
  \caption{\large Regression Results: Impact of renewable energy investments on detected unemployment rate recovery following a change in active mines} 
\begin{tabular}{lcccc}
   \tabularnewline\midrule\midrule
   Dependent Variable: & \multicolumn{4}{c}{$\Delta$Unemployment Rate}\\
   Model:                                                     & (1)             & (2)             & (3)            & (4)\\
   \midrule \emph{Variables} &   &   &   &  \\
   $\Delta$Active Mines\textsubscript{t-2}                                  & 0.0402$^{*}$   & 0.0400$^{*}$   & 0.0433$^{*}$  & 0.0427$^{*}$\\
                                                              & (0.0151)        & (0.0161)        & (0.0175)       & (0.0178)\\
   $\Delta$ Active Mines\textsubscript{t-2} $\times$ REE\textsubscript{t-1} &                 & 0.0049          &                & 0.0229\\
                                                              &                 & (0.0249)        &                & (0.0300)\\
\end{tabular}
\end{table}

\subsection{Spatial Models}

\begin{table}[H]
\caption{\large Impacts of Spatial Error-Lag Model}
\centering
\begin{tabular}{lcccccccc}
\toprule
\multicolumn{9}{c}{Outcome Variable: $\Delta$Unemployment Rate}\\
\cmidrule{3-7}
\multicolumn{1}{c}{} & \multicolumn{2}{c}{$\Delta$ Active Mines\textsubscript{t}} & \multicolumn{2}{c}{$\Delta$ Active Mines\textsubscript{t-1}} & \multicolumn{2}{c}{$\Delta$ Active Mines\textsubscript{t-2}} & \multicolumn{2}{c}{$\Delta$ (log) Real GDPPC\textsubscript{t}} \\
\cmidrule(l{3pt}r{3pt}){2-3} \cmidrule(l{3pt}r{3pt}){4-5} \cmidrule(l{3pt}r{3pt}){6-7} \cmidrule(l{3pt}r{3pt}){8-9}
 & Estimate & Sim. SE & Estimate & Sim. SE & Estimate & Sim. SE & Estimate & Sim. SE\\
\midrule
Direct & -0.036*** & 0.004 & -0.008. & 0.004 & 0.027*** & 0.004 & -0.550*** & 0.027\\
Indirect & -0.215*** & 0.027 & -0.045. & 0.025 & 0.163*** & 0.028 & -3.308*** & 0.191\\
Total & -0.250*** & 0.032 & -0.053. & 0.029 & 0.190*** & 0.032 & -3.858*** & 0.217\\
\bottomrule
\multicolumn{9}{l}{\emph{Signif. Codes: ***: 0.001, **: 0.01, *: 0.05, .: 0.1}}\\
\end{tabular}
\end{table}

\begin{table}[H]
\caption{\large Impacts of Spatial Lag Model}
\centering
\begin{tabular}{lcccccccc}
\toprule
\multicolumn{9}{c}{Outcome Variable: $\Delta$Unemployment Rate}\\
\cmidrule{3-7}
\multicolumn{1}{c}{} & \multicolumn{2}{c}{$\Delta$ Active Mines\textsubscript{t}} & \multicolumn{2}{c}{$\Delta$ Active Mines\textsubscript{t-1}} & \multicolumn{2}{c}{$\Delta$ Active Mines\textsubscript{t-2}} & \multicolumn{2}{c}{$\Delta$ (log) Real GDPPC\textsubscript{t}} \\
\cmidrule(l{3pt}r{3pt}){2-3} \cmidrule(l{3pt}r{3pt}){4-5} \cmidrule(l{3pt}r{3pt}){6-7} \cmidrule(l{3pt}r{3pt}){8-9}
 & Estimate & Sim. SE & Estimate & Sim. SE & Estimate & Sim. SE & Estimate & Sim. SE\\
\midrule
Direct & -0.036*** & 0.005 & -0.009. & 0.005 & 0.028*** & 0.005 & -0.651*** & 0.032\\
Indirect & -0.082*** & 0.012 & -0.020. & 0.011 & 0.064*** & 0.010 & -1.490*** & 0.080\\
Total & -0.118*** & 0.017 & -0.029. & 0.016 & 0.091*** & 0.015 & -2.141*** & 0.111\\
\bottomrule
\multicolumn{9}{l}{\emph{Signif. Codes: ***: 0.001, **: 0.01, *: 0.05, .: 0.1}}\\
\end{tabular}
\end{table}

\begin{table}[H]
\caption{\large Coefficient Estimates of Spatial Error Model}
\centering
\begin{tabular}{ll}
\toprule
~ & $\Delta$Unemployment Rate\\
\cmidrule{1-2}
\emph{Variables} & ~\\
Spatial error parameter ($\rho$) & 0.742***\\
~ & (0.004)\\
$\Delta$ Active Mines\textsubscript{t} & -0.026***\\
~ & (0.004)\\
$\Delta$ Active Mines\textsubscript{t-1} & -0.009*\\
~ & (0.004)\\
$\Delta$ Active Mines\textsubscript{t-2} & 0.021***\\
~ & (0.004)\\
$\Delta$ (log) Real GDPPC\textsubscript{t} & -0.543***\\
~ & (0.03)\\
\bottomrule
\multicolumn{2}{l}{\emph{Signif. Codes: ***: 0.001, **: 0.01, *: 0.05, .: 0.1}}\\
\end{tabular}
\end{table}

\begin{table}[H]
\centering
\caption{\large Spatial Model Comparison by Criterion}
\begin{tabular}{l|r|r|r}
\hline
  &  Log Likelihood & AIC & BIC\\
\hline
Spatial Lag-Error Model & -127,721.6 & 261,635.3 & 289,253.0\\
\hline
Spatial Lag Model & -272,367.1 & 550,924.2 & 578,533.0\\
\hline
Spatial Error Model & -272,426.2 & 551,042.3 & 578,651.1\\
\hline
\end{tabular}
\end{table}

\begin{table}[H]
\centering
\caption{\large Impacts of Spatial Lag-Error (SARAR) models of overall county economy employment indicators}
\begin{tabular}{lcccccc}
\toprule
\multicolumn{1}{c}{} & \multicolumn{3}{c}{$\Delta$ (log) Employed Persons} & \multicolumn{3}{c}{$\Delta$ (log)Unemployed Persons} \\
\cmidrule(l{3pt}r{3pt}){2-4} \cmidrule(l{3pt}r{3pt}){5-7}
  & Direct & Indirect & Total & Direct & Indirect & Total\\
\midrule
$\Delta$ Active Mines\textsubscript{t} & 0.001** & 0.002** & 0.002** & -0.003*** & -0.022*** & -0.026***\\
~ & (0.000) & (0.001) & (0.001) & (0.001) & (0.004) & (0.005)\\
\addlinespace
$\Delta$ Active Mines\textsubscript{t-1} & 0.001*** & 0.003*** & 0.004*** & 0.000 & -0.002 & -0.003\\
~ & (0.000) & (0.000) & (0.001) & (0.001) & (0.004) & (0.004)\\
\addlinespace
$\Delta$ Active Mines\textsubscript{t-2} & 0.000 & 0.000 & 0.001 & 0.003*** & 0.017*** & 0.019***\\
~ & (0.000) & (0.000) & (0.001) & (0.001) & (0.004) & (0.004)\\
\end{tabular}
\end{table}

\subsection{Heterogeneous Trends Models}

\begin{table}[H]
\centering
\caption{\large Coefficient estimates of 1- and 2-factor models}
\begin{tabular}{lcc}
\toprule
& (1)& (2)\\
\multicolumn{1}{c}{} & \multicolumn{2}{c}{$\Delta$Unemployment Rate}\\
\midrule
\emph{Variables} & ~ & ~\\
$\Delta$ Active Mines\textsubscript{t} & -0.065*** & -0.065***\\
~  & (0.006) & (0.007)\\
$\Delta$ Active Mines\textsubscript{t-1} & -0.007 & -0.007\\
~  & (0.006) & (0.006)\\
$\Delta$ Active Mines\textsubscript{t-2} & 0.037*** & 0.037***\\
~  & (0.006) & (0.006)\\
$\Delta$ (log) Real GDPPC\textsubscript{t} & -0.941*** & -0.941***\\
~ & (0.042) & (0.043)\\
\bottomrule
\addlinespace
R$^2$ & 0.384         & 0.193  \\
Additive Effects Type: &  two-ways &  two-ways \\
\# Unobserved Factors Used: & 1 & 2 \\
\bottomrule
\addlinespace
\multicolumn{3}{l}{\emph{Signif. Codes: ***: 0.001, **: 0.01, *: 0.05, .: 0.1}}\\
\end{tabular}
\end{table}

\begin{table}[H]
\centering
\caption{\large Coefficient estimates of 1-and 2- factor models incorporating Renewable Energy Investments}
\begin{tabular}{lcc}
\toprule
~& (1)& (2)\\
\multicolumn{1}{c}{} & \multicolumn{2}{c}{$\Delta$Unemployment Rate}\\
\midrule
$\Delta$ Active Mines\textsubscript{t} & -0.069*** & -0.069***\\
~ & (0.007) & (0.007)\\
$\Delta$ Active Mines\textsubscript{t-1} & -0.007 & -0.007\\
~ & (0.007) & (0.007)\\
$\Delta$ Active Mines\textsubscript{t-2} & 0.040*** & 0.040***\\
~ & (0.006) & (0.007)\\
REE $\geq 0.1$\%  of GDP\textsubscript{t} & -0.057*** & -0.057***\\
~ & (0.013) & (0.014)\\
REE $\geq 0.1$\%  of GDP\textsubscript{t-1} & -0.016 & -0.016\\
~ & (0.014) & (0.014)\\
REE $\geq 0.1$\%  of GDP\textsubscript{t-2} & 0.007 & 0.007\\
~ & (0.015) & (0.015)\\
$\Delta$ Active Mines\textsubscript{t} $\times$ REE & 0.073* & 0.073*\\
~ & (0.029) & (0.030)\\
$\Delta$ Active Mines\textsubscript{t-1} $\times$ REE & 0.030 & 0.030\\
~ & (0.032) & (0.033)\\
$\Delta$ Active Mines\textsubscript{t-2} $\times$ REE & -0.031 & -0.031\\
~ & (0.031) & (0.032)\\
$\Delta$ (log) Real GDPPC\textsubscript{t} & -0.962*** & -0.962***\\
~ & (0.044) & (0.045)\\
\bottomrule
\addlinespace
R$^2$ & 0.434 & 0.434\\
Additive Effects Type & two-ways & two-ways\\
\# Unobserved Factors Used & 1 & 2\\
\bottomrule
\addlinespace
\multicolumn{3}{l}{\emph{Signif. Codes: ***: 0.001, **: 0.01, *: 0.05, .: 0.1}}\\
\end{tabular}
\end{table}

\begin{table}[H]
\centering
\caption{\large Coefficient estimates of 1-factor models of each county employment/economy indicator}
\begin{tabular}{lcccc}
\toprule
\multicolumn{1}{l}{\emph{Variables}} & \multicolumn{1}{c}{$\Delta$ Employed Persons} & \multicolumn{1}{c}{$\Delta$ Unemployed Persons} & \multicolumn{1}{c}{$\Delta$ Labour Force}& \multicolumn{1}{c}{$\Delta$ Population} \\
~& (log)& (log)& (log)& (log)\\
\cmidrule(l{3pt}r{3pt}){1-1} \cmidrule(l{3pt}r{3pt}){2-2} \cmidrule(l{3pt}r{3pt}){3-3} \cmidrule(l{3pt}r{3pt}){4-4} \cmidrule(l{3pt}r{3pt}){5-5}
$\Delta$ Active Mines\textsubscript{t} & 0.001*** & -0.006*** & 0.000. & 0.000\\
~ & (0.000) & (0.001) & (0.000) & (0.000)\\
$\Delta$ Active Mines\textsubscript{t-1} & 0.002*** & 0.000 & 0.002*** & 0.000\\
~ & (0.000) & (0.001) & (0.000) & (0.000)\\
$\Delta$ Active Mines\textsubscript{t-2} & 0.000 & 0.004*** & 0.001*** & 0.000\\
~ & (0.000) & (0.001) & (0.000) & (0.000)\\
$\Delta$ (log) Real GDP & 0.049*** & -0.100*** & 0.038*** & 0.004***\\
~ & (0.002) & (0.006) & (0.002) & (0.001)\\
$\Delta$(log) Population & 0.495*** & 0.775*** & 0.508*** & ~\\
~ & (0.016) & (0.048) & (0.014) & ~\\
\bottomrule
\addlinespace
R$^2$ & 0.246 & 0.323 & 0.217 & 0.501 \\
Additive Effects Type & two-ways & two-ways & two-ways & two-ways\\
\# Unobserved Factors Used & 1 & 1 & 1 & 1\\
\bottomrule
\multicolumn{4}{l}{\emph{Signif. Codes: ***: 0.001, **: 0.01, *: 0.05, .: 0.1}}\\
\end{tabular}
\end{table}


\section{Testing}
\subsection{Cross-sectional dependence tests}
\begin{table}[H] 
\centering 
  \caption{\large Results for Model 1}
\begin{tabular*}{\textwidth}{@{}  
                            @{\extracolsep{\fill}}lcccc}
\\[-1.8ex]\hline 
\hline \\[-1.8ex] 
Test & Test Statistic & P-value & Alternative Hypothesis & Outcome\\
\hline \\[-1.8ex] 
Pesaran CD test& 291.6 & 0*** & Cross-sectional dependence & Validated* \\
Breusch-Pagan LM test& 90,361& 0*** &Cross-sectional dependence & Validated\\
Scaled LM test&600& 0***&Cross-sectional dependence & Validated*\\
Average correlation coefficient&0.7181&N/A&Cross-sectional dependence & Validated*\\
Average absolute correlation coefficient&0.7192&N/A&Cross-sectional dependence & Validated\\
\bottomrule
\end{tabular*}
*: Results confirmed by randomisation-based versions of the relevant test robust to global dependence by common factors and persistence
in the data.\\
\end{table}

\begin{table}[H]
\centering 
  \caption{\large Testing for Spatial Dependence in Residuals of Model 1 \& Models 1\textsuperscript{[SEM, SLM, SARAR]} } 
\begin{tabular*}{\textwidth}{@{}  
                            @{\extracolsep{\fill}}llccc}
\\[-1.8ex]\hline 
\hline \\[-1.8ex] 
Model \# & Method & Test Statistic & P-value & Alternative Hypothesis\\
\hline\\[-1.8ex] 
Model 1 & Pesaran CD test & 43.928 & 0.000  & Cross-sectional dependence\\
~ & Average correlation coefficient & 0.005 & N/A & Cross-sectional dependence\\
\hline\\[-1.8ex] 
Model 1\textsuperscript{SEM} & Pesaran CD test & -2.399 & 0.016  & Cross-sectional dependence\\
~ & Average correlation Coefficient & 0.000 & N/A  & cross-sectional dependence\\
\hline\\[-1.8ex] 
Model 1\textsuperscript{SLM}& Pesaran CD test & -2.098 & 0.036 & Cross-sectional dependence\\
~ & Average correlation coefficient & 0.000 & N/A  & Cross-sectional dependence\\
\hline\\[-1.8ex] 
Model 1\textsuperscript{SARAR} & Pesaran CD test & -2.029 & 0.042 & Cross-sectional dependence\\
~ & Average correlation coefficient & 0.000  & N/A & Cross-sectional dependence\\
\hline \\[-1.8ex] 
\end{tabular*}
Ideally, a Moran’s I test would be implemented to determine the degree of residual spatial dependence in each of the spatial models. Unfortunately, such a method in R only exists for cross-sectional data and not panel data. As an imperfect alternative, we applied Pesaran’s cross-sectional dependence test to the residuals of each spatial model (as well as the residuals of Model 1). This revealed a small degree of remaining spatial dependence. However, we are only able to reject the null hypothesis of no cross-sectional dependence at a 5\% level of significance for each spatial model as compared to a less than 0.01\% level of significance in Model 1 without spatial components. The average correlation coefficient is also near 0 for the residuals of each spatial model, whereas the average correlation coefficient of the residuals of Model 1 is 0.005.\\
\end{table}


\subsection{Optimal factor choice testing for heterogeneous trends models}
\begin{figure}[H]
    \centering
    \caption{\large Scree Plot: Optimal Factor Choices Reported by Various Factor/Principal Component Analysis Methods}
    \includegraphics[width = 16cm]{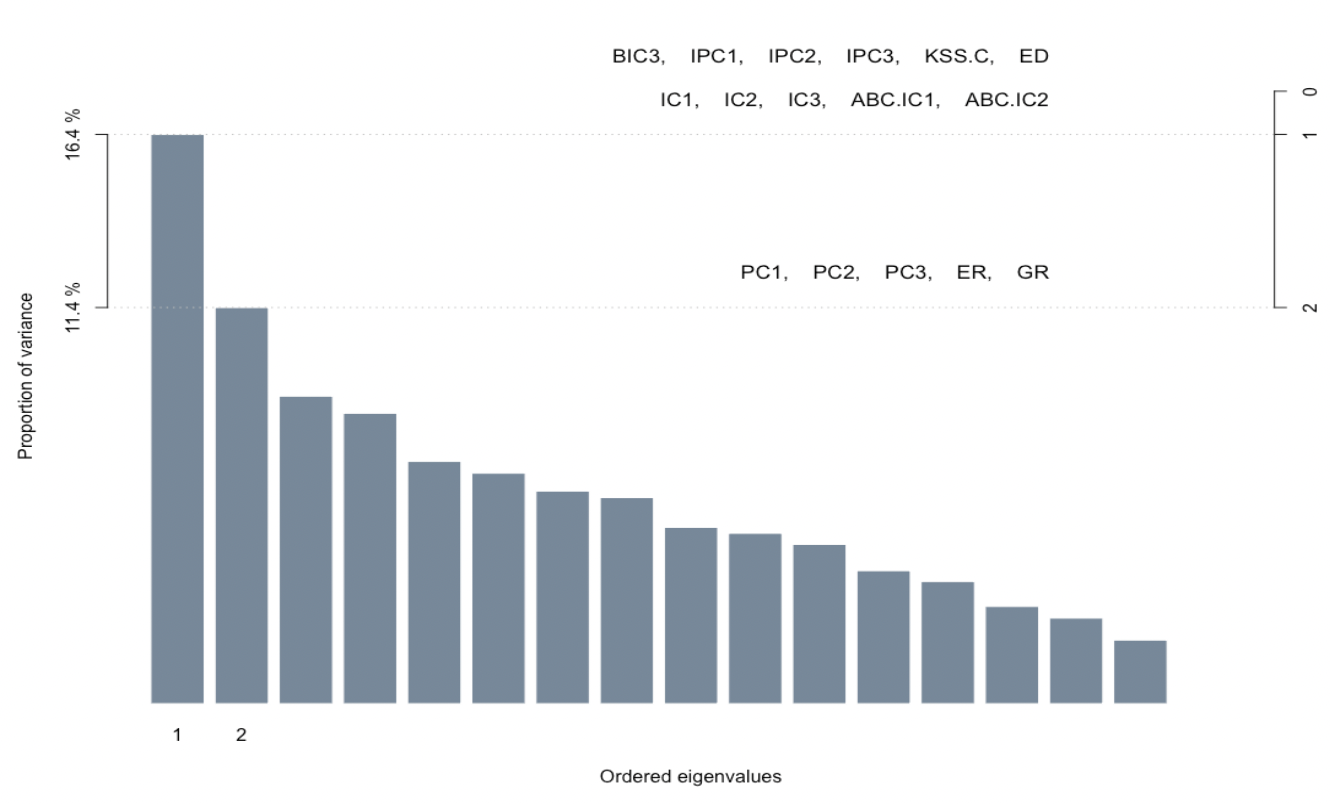}
    \label{fig:screefactordims}
\end{figure}

\subsection{Optimal cluster choice for typology construction}
\begin{figure}[H]
    \centering
    \caption{\large Optimal Cluster (k) Choice: Within Cluster Sum of Squares (Elbow Method)}
    \includegraphics[width = 16cm]{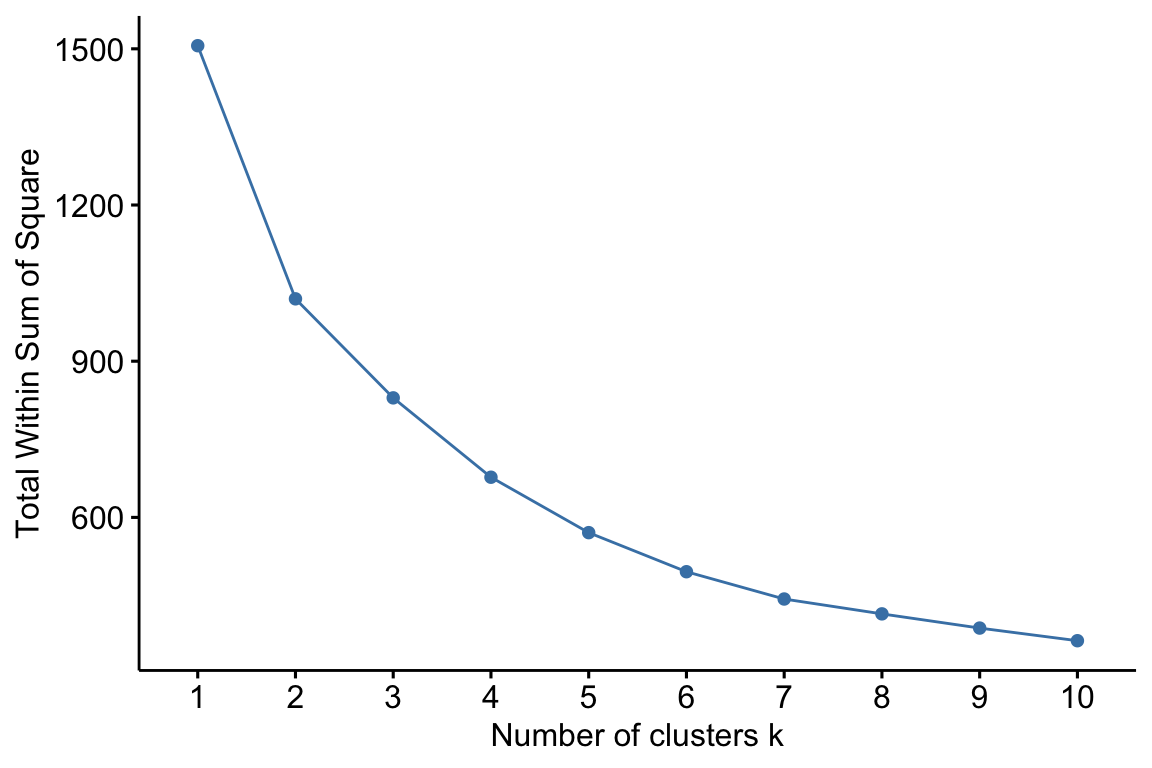}
    \label{fig:elbow}
\end{figure}

\begin{figure}[H]
    \centering
    \caption{\large Optimal Cluster (k) Choice: Average Silhouette Method}
    \includegraphics[width = 16cm]{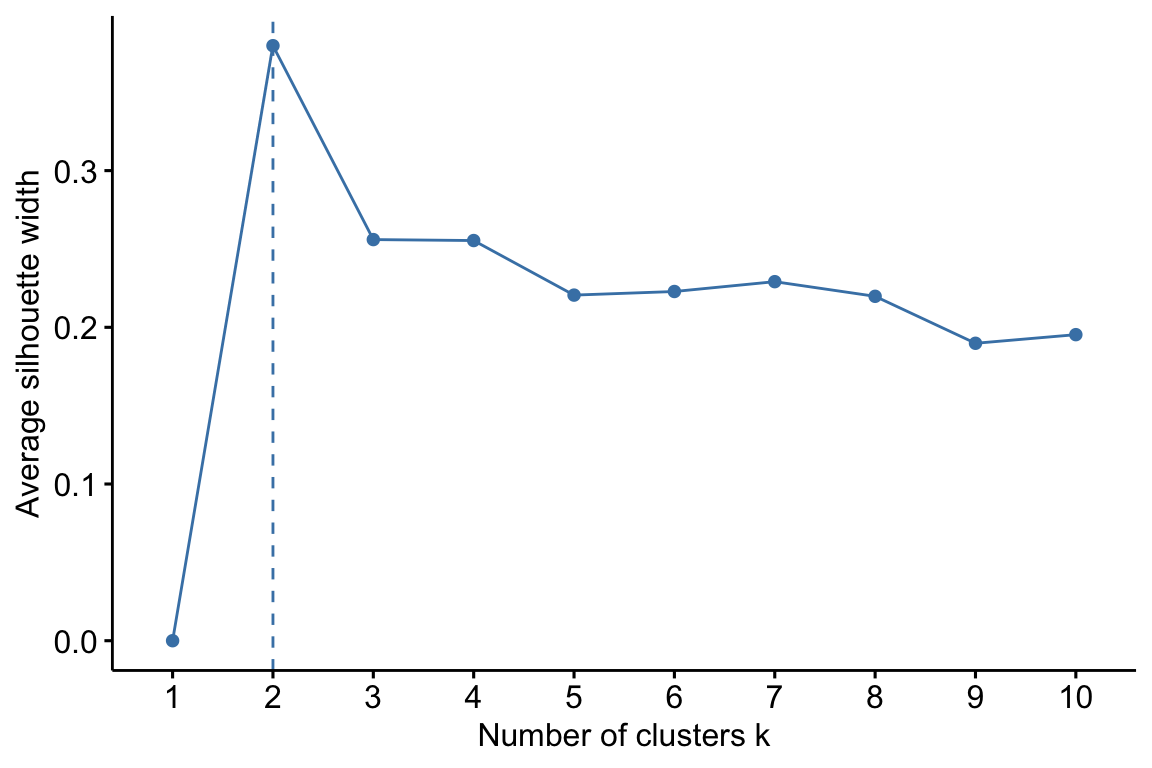}
    \label{fig:silhouette}
\end{figure}

\begin{figure}[H]
    \centering
    \caption{\large Optimal Cluster (k) Choice: Gap Statistic}
    \includegraphics[width = 16cm]{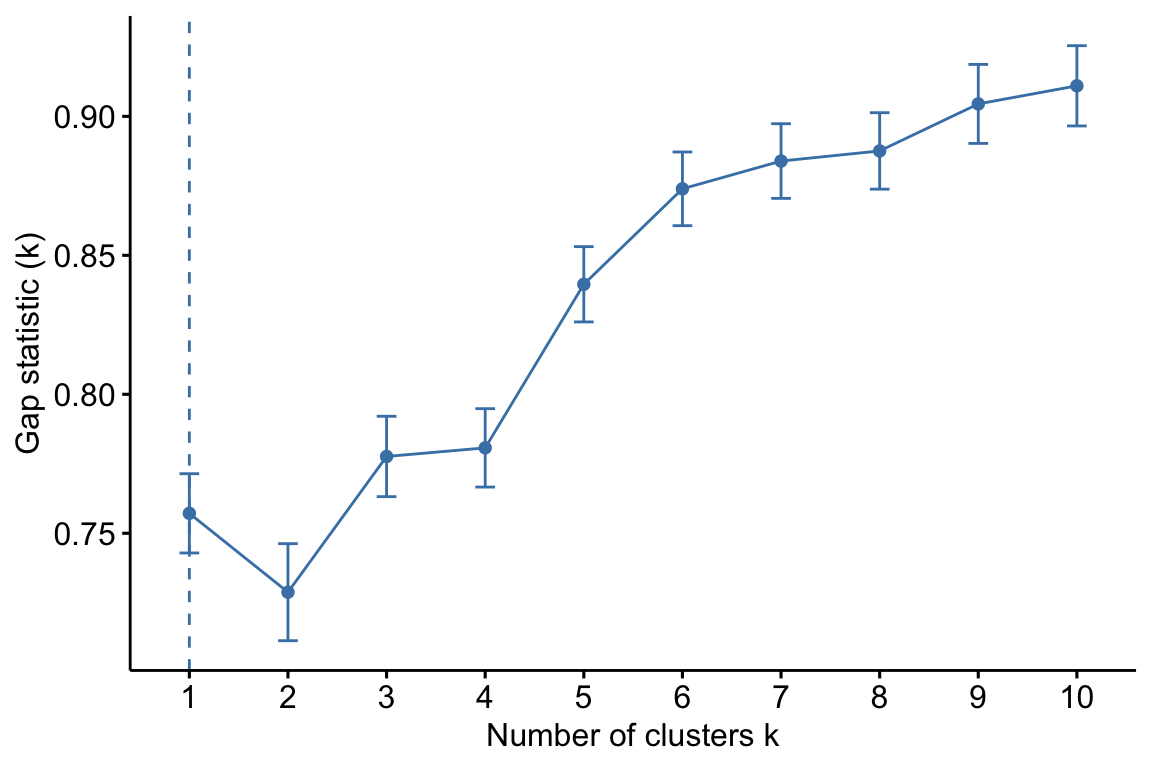}
    \label{fig:gapstat}
\end{figure}

\section{Typology}
\subsection{Typology clustering results}
\begin{table}[H]
\centering
\caption{\large Typology of US Coal Counties}
\begin{tabular}{lcccc}
  \hline
 & \emph{Type 1} & \emph{Type 2} & \emph{Type 3} & \emph{US Average} \\ 
  \hline
  Rural vs. Urban & 2.5 & 5.3 & 7.3 & 5.0 \\ 
  ~& (Metro) & (Nonmetro) & (Rural/Nonmetro) & (Nonmetro) \\ 
      ~& ~ & ~ & ~ & ~\\
  Population & 194,262 & 36,100 & 19,143 & 104,727 \\
    ~& ~ & ~ & ~ & ~\\
  Educational attainment & 29.1\% & 16.5\% & 11.4\% & 21.9\% \\
    ~& ~ & ~ & ~ & ~\\
  Median earnings & \$38,686 & \$33,948 & \$29,845 & \$35,826 \\ 
    ~& ~ & ~ & ~ & ~\\
  Female labor force participation rate & 72.7\% & 66.1\% & 49.2\% & 69.7\% \\ 
    ~& ~ & ~ & ~ & ~\\
  Economic Diversity Index & High/ & Medium/ & Low & Medium \\ 
    ~& Medium & Low & & \\
        ~& ~ & ~ & ~ & ~\\
\# counties that voted for the Republican  &36&156&41&2,568 \\
 Party in the 2020 presidential election &&&\\
    \\[-1em]
    
\# counties that voted for the Republican &36&156&41&2,623 \\
 Party in the 2016 presidential election &&&\\
    \\[-1em]
    
  \emph{Total number of counties per type} & \emph{50} & \emph{160} & \emph{42} & \emph{3,107} \\ 
   \hline
\end{tabular}
\end{table}

\subsection{Incorporating the typology into the econometric models}

\begingroup
\centering
\begin{table}[H]
\centering
\caption{\large Impact of change in active mines on county employment indicators\\
\scriptsize Allowing for slope heterogeneity of treatment variables}
\begin{tabular}{lcc}
   \tabularnewline \midrule \midrule
   Dependent Variable: & \multicolumn{2}{c}{$\Delta$Unemployment Rate}\\
   Model:                                     & (1)            & (2)\\  
   \midrule
   \emph{Variables}\\
  Type 1 $\times$ $\Delta$ Active Mines\textsubscript{t+1}  &                & -0.0141\\   
                                              &                & (0.0133)\\   
   Type 2 $\times$ $\Delta$ Active Mines\textsubscript{t+1}  &                & -0.0184\\   
                                              &                & (0.0121)\\   
   Type 3 $\times$ $\Delta$ Active Mines\textsubscript{t+1}  &                & -0.0181\\   
                                              &                & (0.0256)\\  
   Type 1 $\times$ $\Delta$ Active Mines\textsubscript{t}        & -0.0063        & -0.0088\\   
                                              & (0.0172)       & (0.0175)\\   
   Type 2 $\times$ $\Delta$ Active Mines\textsubscript{t}        & -0.0088        & -0.0113\\   
                                              & (0.0084)       & (0.0092)\\   
   Type 3 $\times$ $\Delta$ Active Mines\textsubscript{t}        & -0.0598$^{**}$ & -0.0609$^{**}$\\   
                                              & (0.0167)       & (0.0174)\\   
   Type 1 $\times$ $\Delta$ Active Mines\textsubscript{t-1}  & -0.0236        & -0.0250\\   
                                              & (0.0154)       & (0.0162)\\   
   Type 2 $\times$ $\Delta$ Active Mines\textsubscript{t-1}  & 0.0016         & 0.0001\\   
                                              & (0.0102)       & (0.0097)\\   
   Type 3 $\times$ $\Delta$ Active Mines\textsubscript{t-1}  & -0.0042        & -0.0049\\   
                                              & (0.0133)       & (0.0151)\\   
   Type 1 $\times$ $\Delta$ Active Mines\textsubscript{t-1}  & -0.0267$^{.}$  & -0.0294$^{*}$\\   
                                              & (0.0129)       & (0.0138)\\   
   Type 2 $\times$ $\Delta$ Active Mines\textsubscript{t-2}  & 0.0242$^{.}$   & 0.0261$^{.}$\\   
                                              & (0.0131)       & (0.0131)\\   
   Type 3 $\times$ $\Delta$ Active Mines\textsubscript{t-2}  & 0.0468$^{***}$ & 0.0473$^{**}$\\   
                                              & (0.0108)       & (0.0126)\\   
 
   Type 1 $\times$ $\Delta$ Active Mines\textsubscript{t-3}  &                & -0.0117\\   
                                              &                & (0.0119)\\   
   Type 2 $\times$ $\Delta$ Active Mines\textsubscript{t-3}  &                & 0.0060\\   
                                              &                & (0.0061)\\   
   Type 3 $\times$ $\Delta$ Active Mines\textsubscript{t-3}  &                & 0.0049\\   
                                              &                & (0.0107)\\   
    $\Delta$ (log) Real GDPPC\textsubscript{t}                 & -1.577$^{***}$ & -1.588$^{***}$\\   
                                              & (0.3474)       & (0.3561)\\                                             
   \midrule
   \emph{Fixed-effects}\\
   County FIPS Code                           & Yes            & Yes\\  
   Year                                       & Yes            & Yes\\  
   \midrule
   \emph{Fit statistics}\\
   Observations                               & 4,518          & 4,267\\  
   R$^2$                                      & 0.65772        & 0.66275\\  
   Within R$^2$                               & 0.05330        & 0.05772\\  
   \midrule \midrule
   \multicolumn{3}{l}{\emph{Clustered (County FIPS Code \& Year) standard-errors in parentheses}}\\
   \multicolumn{3}{l}{\emph{Signif. Codes: ***: 0.001, **: 0.01, *: 0.05, .: 0.1}}\\
\end{tabular}
\end{table}
\par\endgroup

\end{appendices}

\end{document}